\documentclass[aps,twocolumn,amsmath,amssymb,pra,notitlepage,floatfix]{revtex4-1}

\usepackage{graphicx}
\usepackage{dcolumn}
\usepackage{bm}
\usepackage[colorlinks=true,linkcolor=blue]{hyperref}

\newcommand{\eat}[1]{}

\begin{document}

\title{Dynamical quantum phase transitions in $U(1)$ quantum link models}

\author{Yi-Ping Huang}
\author{Debasish Banerjee}
\author{Markus Heyl}
\affiliation{Max-Planck-Institut fur Physik komplexer Systeme, Nothnitzer Str. 38, 01187 Dresden, Germany}
\date{\today}

\begin{abstract}
Quantum link models are extensions of Wilson-type lattice gauge theories which realize exact
gauge invariance with finite-dimensional Hilbert spaces.
Quantum link models not only reproduce the standard features of Wilson's
lattice gauge theories, but also host new phenomena such as crystalline
confined phases.
We study the non-equilibrium quench dynamics for two representative cases, $U(1)$
quantum link models in (1+1)d and (2+1)d,  
through the lens of dynamical quantum
phase transitions.
Finally, we discuss the connection to the high-energy perspective and
the experimental feasibility to observe the discussed phenomena in recent quantum simulator
settings such as trapped ions, ultra-cold atoms, and Rydberg atoms.
\end{abstract}

\maketitle

\begin{figure}[htp]
	\begin{center}
		\includegraphics[width=\columnwidth]{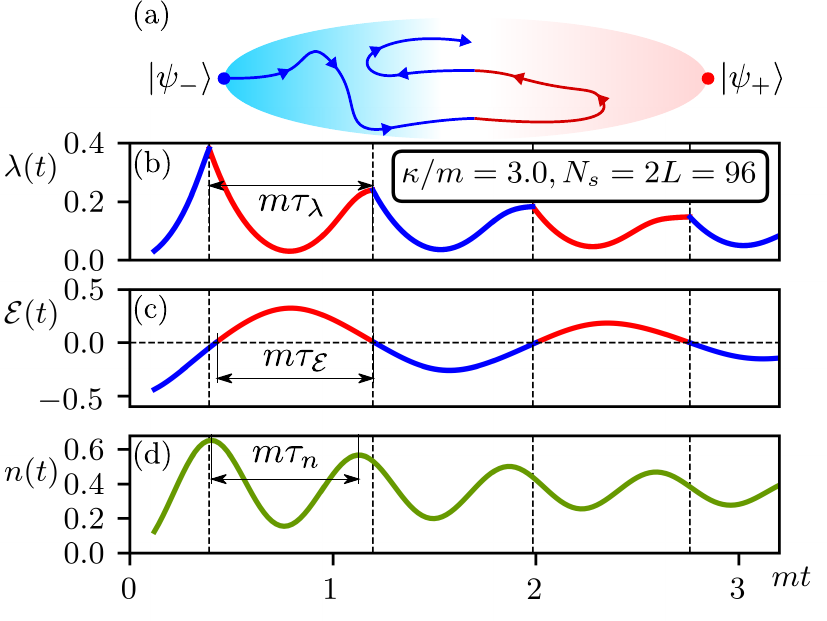}
	\end{center}
	\caption{
	(a) Schematic plot of the wave function dynamics in Hilbert space of the
	considered lattice gauge theories. The two symmetry-broken ground states
	$|\psi_{\pm}\rangle$ represent extremal points, where the order parameter
	takes maximal absolute value. Starting at $|\psi_{-}\rangle$ the state
	explores the Hilbert space. The projection onto the ground state manifold
	classifies the state according to whether the state is closer to
	$|\psi_{-}\rangle$ or $|\psi_{+}\rangle$ (blue or red).
	(b) 
	The dynamics of the dominant rate function $\lambda(t)$ of the full return probability.
	The blue and red colors represent the dominant components $\lambda_{-}(t)$ and $\lambda_{+}(t)$, respectively.  	
	The vertical dashed lines mark the times when $\lambda(t)$ has kinks and
	undergoes a DQPT, switching between the two components.
	We compare $\lambda(t)$ to the dynamics of the order parameter $\mathcal{E}(t) $ (c)  and the fermionic matter particle density $n(t)$ (d).
	}
	\label{fig:1}
\end{figure}

\emph{Introduction -- } 
Gauge theories play an important role in physics ranging from the high-energy
context~\cite{peskin1995quantum}
to models for quantum memories~\cite{Dennis2002} and effective low-energy descriptions for condensed
matter systems~\cite{Lee2006,Lucile2017}.
Today, synthetic quantum systems, such as realized in ultra-cold atoms in optical
lattices and trapped ions, promise to provide a controlled experimental access to
the unitary quantum evolution in lattice gauge theories
(LGTs)~\cite{Wiese2013,Banerjee2013,Hauke2013,Rico2014,Buyens2014,Stannigel2014,Dalmonte2016}, as demonstrated
recently on a digital quantum simulator~\cite{Martinez2016}.
This perspective has stimulated significant interest in the real-time dynamics of
LGTs~\cite{Pichler2016}.
LGTs display various important dynamical phenomena, which are concerned with the
evolution of an initial vacuum subject to a perturbation, such as the Schwinger
mechanism or vacuum decay~\cite{Schwinger1951,Coleman1975charge,Coleman1976more,Kogut2003}.
Recently, it has been observed that the decay of a vacuum in quantum many-body systems
can undergo a dynamical quantum phase transition (DQPT)~\cite{Markus2013,Heyl2018Rev} 
appearing as a real-time non-analytic behavior in the Loschmidt echo or vacuum persistence
probability~\cite{Jurcevic2017,Zhang2017}.
Up to now, it is, however, an open question to which extent also gauge theories can
undergo DQPTs and what the consequence would be for the general physical properties of
such systems.
In this work, we investigate the vacuum dynamics of $U(1)$ lattice gauge
theories exhibiting symmetry-broken phases in equilibrium.
Initializing the system in a symmetry-broken vacuum, we study the real-time evolution
as a consequence of a Hamiltonian perturbation.
Instead of monitoring the full detail of the time-evolved wave
function in many-body Hilbert space, we investigate the dynamics projected
to the ground state manifold, which is equivalent to the vacuum persistence
probability for the case of a unique vacuum.
The information obtained by the projection onto this subspace is illustrated in Fig.~\ref{fig:1}a, where we represent the states in Hilbert space by ordering them according to their order parameter expectation value.
In this picture, the symmetry-broken ground states of the initial Hamiltonian constitute extremal points, 
illustrated here for a broken $\mathbb{Z}_2$ symmetry as studied in this work.
For the more general case $\mathbb{Z}_n$ there will be more of such extremal points accordingly.
Starting in one of the vacua, the time-evolved quantum many-body state traverses through Hilbert space, eventually crossing over to states closer to the other vacuum.
It is the property of the proposed projection onto the vacua subspace to capture the switching between different branches of Hilbert space.
We find that such a switching can occur only in a nonanalytic fashion implying a DQPT in nonequilibrium real-time dynamics.
A signature of the switching and the DQPT can be detected from local observables via the order parameter that has to change sign in the proximity respective point in time.

\emph{Quantum link models --}
Gauge theories are theories with hard local constraints enforced via Gauss Law, and
can be defined non-perturbatively on a lattice~\cite{Kogut2003}. 
Within the Wilson formulation of LGTs, the gauge fields are defined on the bonds
connecting the lattice sites, where matter fields reside.
Quantum link models~(QLMs) extend Wilson's LGT formulation using
finite-dimensional Hilbert spaces for gauge fields~\cite{Chandrasekharan1997,Wiese2013}.
On the one hand, such finite-dimensional Hilbert spaces are often easier to simulate numerically yielding a computational advantage.
On the other hand, such LGTs can also exhibit a host of physical phenomena,
qualitatively different from Wilson's LGT, such as crystalline confined and deconfined
phases~\citep{Moessner2001a,Moessner2001b,Huse2003,Shannon2004,Chakravarty2002,Banerjee2013b,Banerjee2018},
existence of soft modes~\cite{Banerjee2014}, deconfined Rokshar-Kivelson
points~\cite{Rokhsar1988}, and the realization of massless chiral
fermions~\cite{Brower1999}.
In this work, we go beyond the equilibrium context and consider the quench dynamics of $U(1)$ invariant QLMs with spin-$1/2$ quantum links both in $(1+1)$d and $(2+1)$d.
We concentrate on the gauge-field dynamics, which in $(1+1)$d is achieved minimally by a coupling of the $U(1)$ quantum links to matter fields residing on the lattice site, while in $(2+1)$d the gauge fields generate their own dynamics without the need to couple to matter degrees of freedom.
The Hamiltonian for the $(1+1)$d system of size $L$ is:
\begin{eqnarray}
	H_{1D}=-\kappa \sum_{x=0}^{L-2}
	(\psi^{\dagger}_{x}U_{x,\hat{i}}\psi_{x+\hat{i}}+h.c.)
	+\sum_{x=0}^{L-1}m p_x \psi^{\dagger}_{x}\psi_{x},
	\label{1d_QLM}
\end{eqnarray}
where $\psi^{\dagger}_{x}(\psi_{x})$ denotes the matter fermion creation(annihilation) operator on site $x$; 
$m>0$ is the bare mass of the fermions;
$U_{x,\hat{i}}$ is the quantum link operator representing the gauge field on the link 
connecting sites $x$ and $x+\hat{i}$ where $\hat{i}$ is the unit vector for the 1d lattice; 
$p_x=(-1)^x$ and 
$L$ is the number of matter fields where the total number of degrees
of freedom is $N_s=2L$. We define the Hamiltonian using open boundary condition
where the effect on the bulk physics is negligible at the thermodynamic limit.
The Hamiltonian we study for the $(2+1)$d system is: 
\begin{eqnarray}
	H_{2D}&=& \sum_{\Box}-J\left( U_{\Box}+U^{\dagger}_{\Box}
	\right)
	+V \left( U_{\Box}+U^{\dagger}_{\Box}
	\right)^2\text{,}
	\label{2d_QLM}
\end{eqnarray}
where $U_{\Box}=U_{x,\hat{i}}U_{x+\hat{i},\hat{j}}U^{\dagger}_{x+\hat{j},\hat{i}}U^{\dagger}_{x,\hat{j}}$
and $\hat{i},\hat{j}$ denote the unit vectors for the square lattice.

Both Hamiltonians exhibit a $U(1)$ gauge symmetry generated by $G_{x}=\psi^{\dagger}_{x}\psi_{x}-\sum_{\hat{\mu}}\left(
E_{x,\hat{\mu}}-E_{x-\hat{\mu},\hat{\mu}} \right)$ where
$\left[ H,G_x \right]=0$. When the theory is coupled to fermions, the
gauge symmetry is generated by $\tilde{G}_{x}=G_x+\frac{(-1)^x+1}{2}$ and $\left[
H,\tilde{G}_x \right]=0$. The Eq. (\ref{2d_QLM}) has only gauge fields where
$G_x=\sum_{\hat{\mu}}\left( E_{x,\hat{\mu}}-E_{x-\hat{\mu},\hat{\mu}} \right)$.
The gauge-field operator is canonically conjugate to the electric field
operator, i.e., 
$\left[ E_{x,\hat{\mu}},U_{x',\hat{\mu'}}
\right]=U_{x,\hat{\mu}}\delta_{x,x'}\delta_{\mu,\mu'}$ and $\left[
E_{x,\hat{\mu}},U^{\dagger}_{x',\hat{\mu'}}
\right]=-U^{\dagger}_{x,\hat{\mu}}\delta_{x,x'}\delta_{\mu,\mu'}$.
For the spin-$1/2$ QLM model we consider, they are $U_{x,\hat{\mu}}=S^{+}_{x,\hat{\mu}}$,
$U^{\dagger}_{x,\hat{\mu}}=S^{-}_{x,\hat{\mu}}$ and $E_{x,\hat{\mu}}=S^z_{x,\hat{\mu}}$ where
$S^{+}_{x,\hat{\mu}},S^{-}_{x,\hat{\mu}}$ are the spin raising/lowering operators.
In the above equations, we have dropped the electric field energy contribution $H_{E}=g^2\sum_{x,\hat{\mu}} E_{x,\hat{\mu}}^{2}$, which only gives a constant energy offset for the case of the considered spin-$1/2$ quantum links. 
In the following, we will study the real-time quench dynamics of $H_{1D}$ and
$H_{2D}$.
Before discussing the equilibrium phases of the specific models $H_{1D}$ and $H_{2D}$ and our results on the dynamics, we aim to motivate first our study of the vacuum dynamics and its
connection to equilibrium and dynamical quantum phase transitions in general.

\begin{figure}[htp]
	\begin{center}
		\includegraphics[width=\columnwidth]{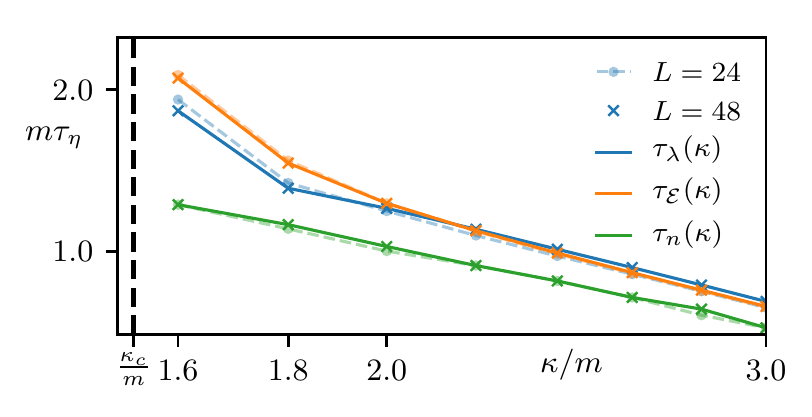}
	\end{center}
	\caption{
	The timescales $\tau_{\lambda}$, $\tau_{\mathcal{E}}$  and
$\tau_{n}$, as defined in Fig.~\ref{fig:1}(b-d), as a function of the coupling $\kappa$
in the final Hamiltonian for two system sizes $L=24,48$.
For $\kappa$ not too close to $\kappa_c$ of the underlying quantum phase transition the timescales $\tau_{\lambda}$ and $\tau_{\mathcal{E}}$ (in contrast to $\tau_n$) are close to each other.
	}
	\label{fig:2}
\end{figure}

\emph{Vacuum dynamics and dynamical quantum phase transitions-- }
As outlined before, we aim to study the dynamics in $U(1)$ QLMs from initial symmetry-broken ground states $|\psi_{\alpha}\rangle,|\psi_{\beta}\rangle$, where $\alpha,\beta = 1, \dots, M$ labels the set of $M$ states in the ground-state manifold.
Motivated by the recent experiment~\cite{Martinez2016}, we choose the initial system parameters such that $|\psi_{\alpha}\rangle$ are product states.
After the quantum quench, the state $|\psi_\alpha(t)\rangle=U(t)|\psi_{\alpha}\rangle$, with $U=e^{-iHt}$ and $H=H_{1D}$ or $H=H_{2D}$, explores the Hilbert space of the
quantum many-body system.
Instead of tracking the full detail of this evolution, we characterize the state's
main properties by the projection onto the ground state manifold of the initial
Hamiltonian via the probabilities $P_{\beta}(t)=|\langle \psi_{\beta}|
\psi_\alpha(t)\rangle|^2$, which, as we will show, provides basic insights to characterize the gauge-field dynamics.
The return probabilities $P_\beta(t)$ also play a central role in the theory of DQPTs~\cite{Heyl2018Rev}.
While equilibrium phase transitions are driven by external control parameters, at DQPTs a system exhibits nonanalytic behavior as a function of time and therefore caused solely by internal dynamics~\cite{Markus2013}.
DQPTs have been initially defined for the case of a unique initial ground state $|\psi_0\rangle$.
It has been a key observation that the return amplitudes $\mathcal{G}(t)=\langle \psi_0 | e^{-iHt} | \psi_0 \rangle$ resemble formally equilibrium partition functions at complex parameters, which have been studied already in the equilibrium case using the concepts of complex partition function zeros~\cite{Lee1952,Yang1952,Fisher1965nature}.
As a consequence, there exists a dynamical counterpart $g(t) =
N_s^{-1}\log[\mathcal{G}(t)]$ to a free energy density, which can become nonanalytic at
critical times. 
While it is important to emphasize that this identification is of formal nature and $g(t)$ is not a thermodynamic potential, it has in the meantime been shown that central properties of equilibrium phase transitions can also be shared by DQPTs.
This includes the robustness against
perturbations~\cite{Karrasch2013,Kriel2014,Markus2015,Sharma2015}, the existence of
order parameters~\cite{Sharma2015,Budich2016,Bhattacharya2017,Flaschner2018,xu2018,guo2018,wang2018,smale2018,tian2018}, or scaling and universality~\cite{Markus2015}.

For the case of degenerate ground state manifolds as we study here, it has turned out
to be very useful to generalize $\mathcal{G}(t)$ to the full return probability $P(t)
= \sum_{\beta=1}^{M} P_\beta(t)
$~\cite{Jurcevic2017,Heyl2014,Weidinger2017,Zunkovic2018,xu2018,guo2018,wang2018,smale2018,tian2018}.
Since $P_{\beta} = e^{-N_s\lambda_{\beta}(t)}$ for $N_s\to\infty$ with $\lambda_\beta(t)$
intensive~\cite{Heyl2014}, there is always a dominant contribution for $P(t)$ in the
sense that for $\lambda(t) = -N_s^{-1}\log[P(t)]$ we have $\lambda(t)\equiv
\min_{\beta}(\left\{ \lambda_{\beta}(t) \right\})$ when $N_s\to\infty$~\cite{Heyl2014}.
Whenever during the dynamics the dominant branch switches from one to the other
vacuum, one obtains a kink in $\lambda(t)$ and therefore a DQPT.
This insight has been used to identify DQPTs in a recent trapped-ion experiment~\cite{Jurcevic2017} will also be utilized here to determine DQPTs in $U(1)$ QLMs in that we compute $\lambda(t)$ via $\lambda(t)= \min_{\beta}(\left\{ \lambda_{\beta}(t) \right\})$.
In the following, we study the vacuum dynamics for $U(1)$ QLMs using numerical methods.
For the $(1+1)$d case, we map the model to a spin model through Jordan-Wigner
transformation and study the problem by means of the time-evolving block decimation (TEBD)
algorithm~\cite{Vidal2003,Schollwock2011,itensor} and for the $(2+1)$d case, we use a Lanczos-based exact diagonalization (ED)~\cite{sandvik2010}.

\begin{figure}[htp]
	\begin{center}
		\includegraphics[width=\columnwidth]{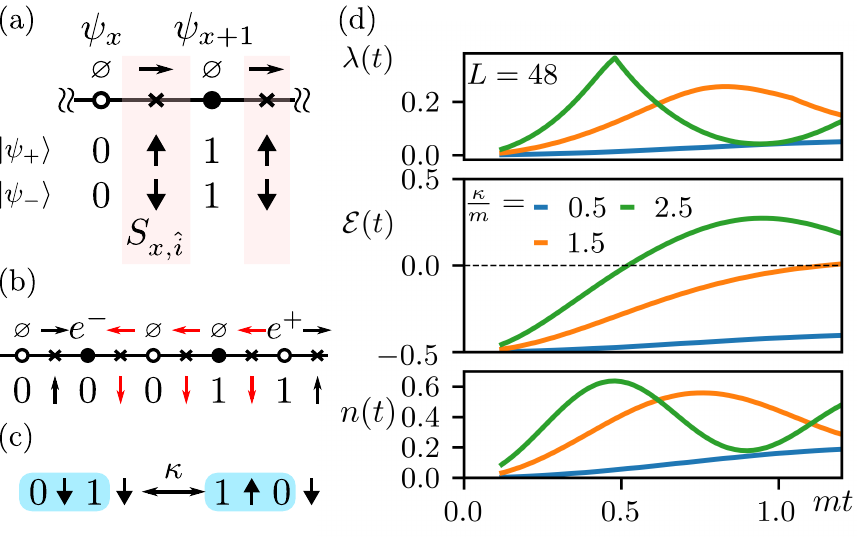}
	\end{center}
	\caption{
		(a) Illustration of the initial product state for the $(1+1)$d quantum link model at $\kappa=0$. 
		The filled/hollow circles are the particle/antiparticle
		field sites, the crosses are the gauge field sites, separated
                by color blocks. 
		The rows with horizontal arrows are the fermionic representation
		of the state. 
		The two symmetry breaking ground states $|\psi_{\pm}\rangle$ are shown. 
		The lower three rows are the wave function representation using spin states.
		(b) Particle-antiparticle pairs can propagate upon flipping the intermediate
		gauge spins.
		(c) The matter-gauge link interaction couples two shaded blocks.
		(d) The dynamics of $\lambda(t)$, $\mathcal{E}(t)$, and $n(t)$ upon changing $\kappa$.
	}
	\label{fig:3}
\end{figure}

\emph{(1+1)d $U(1)$ QLM and quench protocol--}
In equilibrium, the model in Eq. (\ref{1d_QLM}) exhibits a quantum phase transition at
$\kappa_{c}=(0.655)^{-1}m=1.526m$ within the 2d Ising universality class separating a
symmetry-broken phase ($\kappa<\kappa_c$) from a paramagnetic one ($\kappa>\kappa_c$)
with the order parameter $\mathcal{E}=L^{-1}\sum_{x}\langle S^z_{x,\hat{i}}
\rangle$~\cite{Byrnes2002,Rico2014}.
In the following, we study the quench dynamics in this model. In (1+1)d, the staggered
fermions can be converted to local spin operators through Jordan-Wigner
transformation.
We prepare the system initially at $\kappa=0$ in one of the two ground states
$|\psi_{\pm}\rangle = |\pm\rangle_S \otimes |0101 \dots \rangle_\psi$ with
$|0101\dots\rangle_\psi$ the bare vacuum for the matter degrees of
freedom without any particle, see also Fig. \ref{fig:3} (a), and $|+\rangle_S =
|\uparrow \dots\uparrow \rangle$, $|-\rangle_S=|\downarrow \dots \downarrow \rangle$ denoting the fully polarized states for the gauge fields on the links.
Without loss of generality, we start from $|\psi_{-}\rangle$.
When $t>0$, we suddenly turn on $\kappa>0$ and monitor $\lambda(t)$,
$\mathcal{E}(t)$ and the matter particle density $
n(t)\equiv L^{-1}\sum_{x=0}^{L-1}(-1)^x \psi_x^\dag
\psi_x+0.5$.
In $(1+1)d$, the latter can also be identified with the chiral condensate, which, however, is specific to our model and doesn't hold for other systems.
We calculate $\lambda(t)$ via $\lambda(t)=\min(\lambda_{+}(t),\lambda_{-}(t))$ where
$\lambda_{\pm}(t)=-L^{-1} \log[P_{\pm}(t)]$ and $P_\pm(t) = |\langle
\phi_\pm|\psi_{-}(t)|^2$ with $|\phi_{-}\rangle =|\psi_{-}\rangle$ and
$|\phi_{+}\rangle= \psi_0^\dag
\psi_{L-1}U^{\dagger}_{L-1,\hat{i}}|\psi_+\rangle$
Using the states $|\phi_\pm\rangle$ instead of $|\psi_\pm\rangle$ is necessary
because we use  for our numerics open boundary conditions, where $|\psi_\pm\rangle$
are dynamically decoupled.
However, they can be almost transformed into each other up to one particle-hole excitation at the two edges of the chain, which is accounted for by defining $|\phi_{\pm}\rangle$.
For details see~\cite{SupMat}.

In Fig.~\ref{fig:1} we show numerically obtained data for a quench across the underlying quantum phase transition to $\kappa=3m>\kappa_c$.
We observe DQPTs in $\lambda(t)$ at a series of critical times in the form of kinks caused by a crossing of the two rate functions $\lambda_{+}(t)$ and $\lambda_{-}(t)$.
Thus, the DQPTs mark those points in time where the time-evolved state $|\psi_{-}(t)\rangle$ switches between being closer to $|\psi_{+}\rangle$ to being closer to $|\psi_{-}\rangle$ and vice versa.
With this one can classify $|\psi_{-}(t)\rangle$ with either positive or negative order parameter values, respectively, see the general perspective provided in Fig.~\ref{fig:1}(a).
This information allows us to conclude that $\mathcal{E}(t)$ has to change sign across a DQPT.
Accordingly, $\mathcal{E}(t)$ develops an oscillatory behavior for a sequence of such DQPTs, as we find indeed for our numerical data, see Fig.~\ref{fig:1}(c).
Thus, from the projected vacuum dynamics we obtain useful information of the time evolution of the full quantum many-body state. 
While the aforementioned classification of the state allows us to conclude on the order parameter dynamics, the particle density $n(t)$ behaves differently, as we will study in more detail below.
In this way, $n(t)$ and $\lambda(t)$ are not directly correlated to each other, which is different from the from the Schwinger mechanism of particle-antiparticle production where the vacuum persistence probability is directly linked to $n(t)$~\cite{Schwinger1951}.
Even though Gauss law bridges the dynamics of matter field and gauge field,
it does not give a direct link between $\mathcal{E}(t)$ and $n(t)$, but rather gives
$n(t)=2L^{-1}\sum_{x=0}^{L-1}(-1)^xE_{x,\hat{i}}(t)$ up to constant terms.

In order to make the connection between the studied observables quantitative, we compare the timescales $\tau_{\lambda},\tau_{\mathcal{E}}$ and $\tau_{n}$ defined in Fig.~\ref{fig:1} as a function of $\kappa$ for $\kappa>\kappa_c$, which we show in Fig.~\ref{fig:2}.
For $\kappa$ not too close to $\kappa_c$ we find that $\tau_{\lambda}\approx \tau_{\mathcal{E}}$ whereas the oscillation period for $\tau_n< \tau_{\lambda},\tau_{\mathcal{E}}$ is different.
Upon approaching $\kappa_c$ we observe the increasing influence of finite-size effects and that $\tau_{\lambda}$ and $\tau_{\mathcal{E}}$ start to deviate from each other.
We attribute this difference to the way we numerically obtain these timescales.
Ideally, these periods would have been obtained by studying the oscillations for many cycles by performing a Fourier analysis.
Especially close to the equilibrium critical point, however, the involved time scales become large which allows us to reach reliably only the first two DQPTs.
As a consequence, we estimate the oscillation frequency consistently over the full set
of $\kappa$ by the time difference between the second and first DQPT.
Analogously, we define $\tau_{\mathcal{E}}$ and $\tau_{n}$ as the time between the first two zeros of $\mathcal{E}(t)$ and first two maxima of $n(t)$, respectively.

When we lower $\kappa<\kappa_{c}$, the rate function does not develop a singularity anymore and no DQPT is observed within the time window of simulation, see Fig.~\ref{fig:3}(d).
Following the picture in Fig.~\ref{fig:1}(a) we conclude that the system never switches to the other symmetry-broken sector and therefore does not reach the other basin of states related by the $Z_2$ symmetry. 
As a consequence, the order parameter does not change sign during dynamics.

\emph{$(2+1)$d $U(1)$ QLM-}
After analyzing the $(1+1)d$ QLM, we now go one step further to the $(2+1)$d case, where the gauge fields can generate non-trivial dynamics without coupling to matter fields, see Eq.~(\ref{2d_QLM}).
The first term in Eq.~(\ref{2d_QLM}) describes quantum tunneling between configurations, which satisfy the
Gauss law, flipping all spins on the given plaquette.
The second term is the potential term which prefers to maximize the number of
flippable plaquettes.
Recent studies of this model show new types of crystalline confined
phases in equilibrium~\cite{Shannon2004,Banerjee2013b}.
For $V<V_c \approx -0.38$, the ground states, $|\psi_{\pm}\rangle$, spontaneously break the lattice translation,
$T$, as well as charge conjugation, $\mathcal{C}$. For $V=-\infty$ the ground
states are product states of maximum number of flippable plaquettes with different
chirality of $E$ fields~\cite{SupMat}.
At $V=V_c$, the system undergoes a weak first-order transition into a 
phase which only breaks the $T$ symmetry by one lattice spacing.

\begin{figure}[htp]
	\begin{center}
		\includegraphics[width=8.5cm]{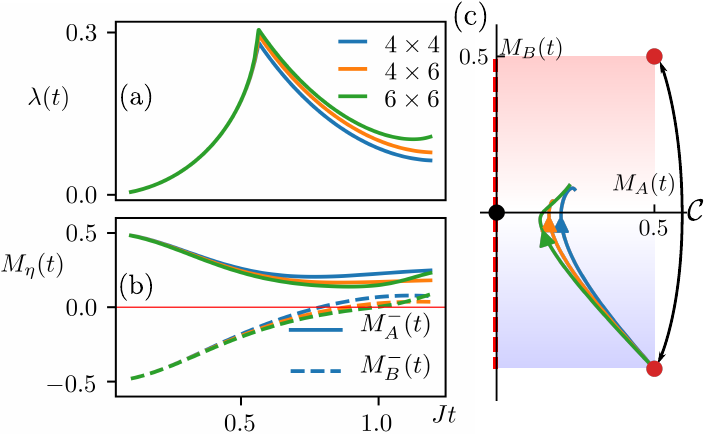}
	\end{center}
	\caption{
		DQPT in the (2+1)d $U(1)$ QLM. The dimensionless time, $t$, uses
		$J^{-1}$ as time scale.
	The quench dynamics using Lanczos-based ED for system size
	$4\times 4$~(32 spins), $4\times 6$~(48 spins) and $6\times 6$~(72 spins).
	(a) The dynamics of the rate function after quench.
	(b) The dynamics of the two-component order parameter defined as the
	height patterns of the plaquettes in the two sublattices, A and B, of the square
        lattice.
	(c) The dynamics of the order parameter after quench in the $(M_A,M_B)$ plane.
	The two red points represent the order related by charge conjugation
	symmetry, $\mathcal{C}$ which transform $M_B\to-M_B$\cite{Banerjee2013b}. 
	Under a lattice translation, the order parameter gets
        reflected about the red dashed line. 
	}
	\label{fig:4}
\end{figure}

We study a quench process similar to $(1+1)d$ case.
For $t\le 0$, we choose $V=-\infty$ such that our system is prepared in a product states, $|\psi_-\rangle$, say. 
When $t>0$, we switch $V$ to $V=0$ and evolve the state using $U_f(t)=e^{-iH_{2D}t}$.
In Fig. \ref{fig:4}(a)  we monitor $\lambda(t)$ and the order parameter, which now exhibits two components  $M_\eta(t)$ with $\eta=A,B$. 
We observe a decisive signature of a DQPT in $\lambda(t)$ with only a weak drift for increasing system size.
This DQPT again marks the sharp transition between the two branches illustrated in Fig.~\ref{fig:1}(a).
Accordingly, we expect a crossover of the order parameter $M_B(t)$ associated with the classification, vanishing at the critical time.
While finite-size effects are still substantial for the individual $M_\eta(t)$, one can observe in Fig.~\ref{fig:4}(c) that the crossing to the other branch in the two-component order parameter plane remains stable upon increasing system size.
From Fig.~\ref{fig:4}(c) we can conclude that  the two-component order parameter 
can move through the path joining $\mathcal{C}$ partners, but not the $T$ partners.
On the studied time scales, $M_A(t)$ therefore does not melt suggesting that the system does not yet restore the lattice translation symmetry. 

\emph{Conclusion-}
The study of dynamics of the lattice gauge theories holds the potential to shed light on many of the
dynamical properties encountered in the phenomenology of high energy physics and of the early
universe. 
The quenched dynamics of the chiral condensate of the Schwinger model can perhaps be
used to qualitatively model the analogous behavior of the condensate of the QCD vacuum
in strong external magnetic fields~\cite{braguta2012,Bali2012} which might occur in heavy-ion collisions
or might have influenced structure formation in the early universe.
The Hamiltonian evolution of the symmetry broken ground state in (2+1)d guided by the
unbroken symmetries is another aspect which might be  argued to hold beyond the
systems considered here and help understand the dynamics of confining theories in nature.
The Hamiltonian with gauge symmetry has been proposed to be realized in quantum
simulators~\cite{Wiese2013,zohar2015} such as quantum circuits~\cite{marcos2014}
, trapped ions~\cite{Hauke2013}, ultracold atoms~\cite{Zohar2011} and Rydberg
atoms~\cite{tagliacozzo2013}.
Initial product state wave function can be prepared with high fidelity
~\cite{lanyon2011,barends2016,Jurcevic2017}. 
To observe the dynamics of the order parameters and particle density, the experiments
need to have local addressability which is possible for quantum simulators such as
trapped ion and Rydberg atoms experiments.
The observation of kinks of Loschmidt echo is challenging, which, however, has been
mastered in a recent trapped ion experiment~\cite{Jurcevic2017},
the photonic quantum walks system~\cite{wang2018,xu2018} and superconducting qubit
simulation~\cite{guo2018}. 
%

%
\emph{Acknowledgment} 
We acknowledge Pochung Chen, Chia-Min Chung, C. -J. David Lin, Ying-Jer Kao, Marcello
Dalmonte and Uwe-Jens Wiese 
for the fruitful discussions.
M. H.  acknowledges
support  by  the  Deutsche Forschungsgemeinschaft  via  the  Gottfried Wilhelm Leibniz
Prize program.

\textbf{Note}:
For a related work on dynamical quantum phase transitions in gauge theories see the article published on the arxiv on the same day by Torsten Zache, Niklas Mueller, Jan Schneider, Fred Jendrzejewski, Jürgen Berges, and Philipp Hauke.

\bibliography{u1QLM_DQPT.bib}

\begin{thebibliography}{65}%
\makeatletter
\providecommand \@ifxundefined [1]{%
 \@ifx{#1\undefined}
}%
\providecommand \@ifnum [1]{%
 \ifnum #1\expandafter \@firstoftwo
 \else \expandafter \@secondoftwo
 \fi
}%
\providecommand \@ifx [1]{%
 \ifx #1\expandafter \@firstoftwo
 \else \expandafter \@secondoftwo
 \fi
}%
\providecommand \natexlab [1]{#1}%
\providecommand \enquote  [1]{``#1''}%
\providecommand \bibnamefont  [1]{#1}%
\providecommand \bibfnamefont [1]{#1}%
\providecommand \citenamefont [1]{#1}%
\providecommand \href@noop [0]{\@secondoftwo}%
\providecommand \href [0]{\begingroup \@sanitize@url \@href}%
\providecommand \@href[1]{\@@startlink{#1}\@@href}%
\providecommand \@@href[1]{\endgroup#1\@@endlink}%
\providecommand \@sanitize@url [0]{\catcode `\\12\catcode `\$12\catcode
  `\&12\catcode `\#12\catcode `\^12\catcode `\_12\catcode `\%12\relax}%
\providecommand \@@startlink[1]{}%
\providecommand \@@endlink[0]{}%
\providecommand \url  [0]{\begingroup\@sanitize@url \@url }%
\providecommand \@url [1]{\endgroup\@href {#1}{\urlprefix }}%
\providecommand \urlprefix  [0]{URL }%
\providecommand \Eprint [0]{\href }%
\providecommand \doibase [0]{http://dx.doi.org/}%
\providecommand \selectlanguage [0]{\@gobble}%
\providecommand \bibinfo  [0]{\@secondoftwo}%
\providecommand \bibfield  [0]{\@secondoftwo}%
\providecommand \translation [1]{[#1]}%
\providecommand \BibitemOpen [0]{}%
\providecommand \bibitemStop [0]{}%
\providecommand \bibitemNoStop [0]{.\EOS\space}%
\providecommand \EOS [0]{\spacefactor3000\relax}%
\providecommand \BibitemShut  [1]{\csname bibitem#1\endcsname}%
\let\auto@bib@innerbib\@empty
\bibitem [{\citenamefont {Peskin}\ and\ \citenamefont
  {Schroeder}(1995)}]{peskin1995quantum}%
  \BibitemOpen
  \bibfield  {author} {\bibinfo {author} {\bibfnamefont {M.~E.}\ \bibnamefont
  {Peskin}}\ and\ \bibinfo {author} {\bibfnamefont {D.~V.}\ \bibnamefont
  {Schroeder}},\ }\href@noop {} {\bibfield  {journal} {\bibinfo  {journal} {The
  Advanced Book Program, Perseus Books Reading, Massachusetts}\ ,\ \bibinfo
  {pages} {4}} (\bibinfo {year} {1995})}\BibitemShut {NoStop}%
\bibitem [{\citenamefont {Dennis}\ \emph {et~al.}(2002)\citenamefont {Dennis},
  \citenamefont {Kitaev}, \citenamefont {Landahl},\ and\ \citenamefont
  {Preskill}}]{Dennis2002}%
  \BibitemOpen
  \bibfield  {author} {\bibinfo {author} {\bibfnamefont {E.}~\bibnamefont
  {Dennis}}, \bibinfo {author} {\bibfnamefont {A.}~\bibnamefont {Kitaev}},
  \bibinfo {author} {\bibfnamefont {A.}~\bibnamefont {Landahl}}, \ and\
  \bibinfo {author} {\bibfnamefont {J.}~\bibnamefont {Preskill}},\ }\href
  {https://aip.scitation.org/doi/10.1063/1.1499754} {\bibfield  {journal}
  {\bibinfo  {journal} {Journal of Mathematical Physics}\ }\textbf {\bibinfo
  {volume} {43}},\ \bibinfo {pages} {4452} (\bibinfo {year}
  {2002})}\BibitemShut {NoStop}%
\bibitem [{\citenamefont {Lee}\ \emph {et~al.}(2006)\citenamefont {Lee},
  \citenamefont {Nagaosa},\ and\ \citenamefont {Wen}}]{Lee2006}%
  \BibitemOpen
  \bibfield  {author} {\bibinfo {author} {\bibfnamefont {P.~A.}\ \bibnamefont
  {Lee}}, \bibinfo {author} {\bibfnamefont {N.}~\bibnamefont {Nagaosa}}, \ and\
  \bibinfo {author} {\bibfnamefont {X.-G.}\ \bibnamefont {Wen}},\ }\href
  {\doibase 10.1103/RevModPhys.78.17} {\bibfield  {journal} {\bibinfo
  {journal} {Rev. Mod. Phys.}\ }\textbf {\bibinfo {volume} {78}},\ \bibinfo
  {pages} {17} (\bibinfo {year} {2006})}\BibitemShut {NoStop}%
\bibitem [{\citenamefont {Savary}\ and\ \citenamefont
  {Balents}(2017)}]{Lucile2017}%
  \BibitemOpen
  \bibfield  {author} {\bibinfo {author} {\bibfnamefont {L.}~\bibnamefont
  {Savary}}\ and\ \bibinfo {author} {\bibfnamefont {L.}~\bibnamefont
  {Balents}},\ }\href {http://stacks.iop.org/0034-4885/80/i=1/a=016502}
  {\bibfield  {journal} {\bibinfo  {journal} {Reports on Progress in Physics}\
  }\textbf {\bibinfo {volume} {80}},\ \bibinfo {pages} {016502} (\bibinfo
  {year} {2017})}\BibitemShut {NoStop}%
\bibitem [{\citenamefont {Wiese}(2013)}]{Wiese2013}%
  \BibitemOpen
  \bibfield  {author} {\bibinfo {author} {\bibfnamefont {U.-J.}\ \bibnamefont
  {Wiese}},\ }\href {\doibase 10.1002/andp.201300104} {\bibfield  {journal}
  {\bibinfo  {journal} {Annalen der Physik}\ }\textbf {\bibinfo {volume}
  {525}},\ \bibinfo {pages} {777} (\bibinfo {year} {2013})}\BibitemShut
  {NoStop}%
\bibitem [{\citenamefont {Banerjee}\ \emph
  {et~al.}(2013{\natexlab{a}})\citenamefont {Banerjee}, \citenamefont
  {B\"ogli}, \citenamefont {Dalmonte}, \citenamefont {Rico}, \citenamefont
  {Stebler}, \citenamefont {Wiese},\ and\ \citenamefont
  {Zoller}}]{Banerjee2013}%
  \BibitemOpen
  \bibfield  {author} {\bibinfo {author} {\bibfnamefont {D.}~\bibnamefont
  {Banerjee}}, \bibinfo {author} {\bibfnamefont {M.}~\bibnamefont {B\"ogli}},
  \bibinfo {author} {\bibfnamefont {M.}~\bibnamefont {Dalmonte}}, \bibinfo
  {author} {\bibfnamefont {E.}~\bibnamefont {Rico}}, \bibinfo {author}
  {\bibfnamefont {P.}~\bibnamefont {Stebler}}, \bibinfo {author} {\bibfnamefont
  {U.-J.}\ \bibnamefont {Wiese}}, \ and\ \bibinfo {author} {\bibfnamefont
  {P.}~\bibnamefont {Zoller}},\ }\href {\doibase
  10.1103/PhysRevLett.110.125303} {\bibfield  {journal} {\bibinfo  {journal}
  {Phys. Rev. Lett.}\ }\textbf {\bibinfo {volume} {110}},\ \bibinfo {pages}
  {125303} (\bibinfo {year} {2013}{\natexlab{a}})}\BibitemShut {NoStop}%
\bibitem [{\citenamefont {Hauke}\ \emph {et~al.}(2013)\citenamefont {Hauke},
  \citenamefont {Marcos}, \citenamefont {Dalmonte},\ and\ \citenamefont
  {Zoller}}]{Hauke2013}%
  \BibitemOpen
  \bibfield  {author} {\bibinfo {author} {\bibfnamefont {P.}~\bibnamefont
  {Hauke}}, \bibinfo {author} {\bibfnamefont {D.}~\bibnamefont {Marcos}},
  \bibinfo {author} {\bibfnamefont {M.}~\bibnamefont {Dalmonte}}, \ and\
  \bibinfo {author} {\bibfnamefont {P.}~\bibnamefont {Zoller}},\ }\href
  {\doibase 10.1103/PhysRevX.3.041018} {\bibfield  {journal} {\bibinfo
  {journal} {Phys. Rev. X}\ }\textbf {\bibinfo {volume} {3}},\ \bibinfo {pages}
  {041018} (\bibinfo {year} {2013})}\BibitemShut {NoStop}%
\bibitem [{\citenamefont {Rico}\ \emph {et~al.}(2014)\citenamefont {Rico},
  \citenamefont {Pichler}, \citenamefont {Dalmonte}, \citenamefont {Zoller},\
  and\ \citenamefont {Montangero}}]{Rico2014}%
  \BibitemOpen
  \bibfield  {author} {\bibinfo {author} {\bibfnamefont {E.}~\bibnamefont
  {Rico}}, \bibinfo {author} {\bibfnamefont {T.}~\bibnamefont {Pichler}},
  \bibinfo {author} {\bibfnamefont {M.}~\bibnamefont {Dalmonte}}, \bibinfo
  {author} {\bibfnamefont {P.}~\bibnamefont {Zoller}}, \ and\ \bibinfo {author}
  {\bibfnamefont {S.}~\bibnamefont {Montangero}},\ }\href {\doibase
  10.1103/PhysRevLett.112.201601} {\bibfield  {journal} {\bibinfo  {journal}
  {Phys. Rev. Lett.}\ }\textbf {\bibinfo {volume} {112}},\ \bibinfo {pages}
  {201601} (\bibinfo {year} {2014})}\BibitemShut {NoStop}%
\bibitem [{\citenamefont {Buyens}\ \emph {et~al.}(2014)\citenamefont {Buyens},
  \citenamefont {Haegeman}, \citenamefont {Van~Acoleyen}, \citenamefont
  {Verschelde},\ and\ \citenamefont {Verstraete}}]{Buyens2014}%
  \BibitemOpen
  \bibfield  {author} {\bibinfo {author} {\bibfnamefont {B.}~\bibnamefont
  {Buyens}}, \bibinfo {author} {\bibfnamefont {J.}~\bibnamefont {Haegeman}},
  \bibinfo {author} {\bibfnamefont {K.}~\bibnamefont {Van~Acoleyen}}, \bibinfo
  {author} {\bibfnamefont {H.}~\bibnamefont {Verschelde}}, \ and\ \bibinfo
  {author} {\bibfnamefont {F.}~\bibnamefont {Verstraete}},\ }\href {\doibase
  10.1103/PhysRevLett.113.091601} {\bibfield  {journal} {\bibinfo  {journal}
  {Phys. Rev. Lett.}\ }\textbf {\bibinfo {volume} {113}},\ \bibinfo {pages}
  {091601} (\bibinfo {year} {2014})}\BibitemShut {NoStop}%
\bibitem [{\citenamefont {Stannigel}\ \emph {et~al.}(2014)\citenamefont
  {Stannigel}, \citenamefont {Hauke}, \citenamefont {Marcos}, \citenamefont
  {Hafezi}, \citenamefont {Diehl}, \citenamefont {Dalmonte},\ and\
  \citenamefont {Zoller}}]{Stannigel2014}%
  \BibitemOpen
  \bibfield  {author} {\bibinfo {author} {\bibfnamefont {K.}~\bibnamefont
  {Stannigel}}, \bibinfo {author} {\bibfnamefont {P.}~\bibnamefont {Hauke}},
  \bibinfo {author} {\bibfnamefont {D.}~\bibnamefont {Marcos}}, \bibinfo
  {author} {\bibfnamefont {M.}~\bibnamefont {Hafezi}}, \bibinfo {author}
  {\bibfnamefont {S.}~\bibnamefont {Diehl}}, \bibinfo {author} {\bibfnamefont
  {M.}~\bibnamefont {Dalmonte}}, \ and\ \bibinfo {author} {\bibfnamefont
  {P.}~\bibnamefont {Zoller}},\ }\href {\doibase
  10.1103/PhysRevLett.112.120406} {\bibfield  {journal} {\bibinfo  {journal}
  {Phys. Rev. Lett.}\ }\textbf {\bibinfo {volume} {112}},\ \bibinfo {pages}
  {120406} (\bibinfo {year} {2014})}\BibitemShut {NoStop}%
\bibitem [{\citenamefont {Dalmonte}\ and\ \citenamefont
  {Montangero}(2016)}]{Dalmonte2016}%
  \BibitemOpen
  \bibfield  {author} {\bibinfo {author} {\bibfnamefont {M.}~\bibnamefont
  {Dalmonte}}\ and\ \bibinfo {author} {\bibfnamefont {S.}~\bibnamefont
  {Montangero}},\ }\href {\doibase 10.1080/00107514.2016.1151199} {\bibfield
  {journal} {\bibinfo  {journal} {Contemporary Physics}\ }\textbf {\bibinfo
  {volume} {57}},\ \bibinfo {pages} {388} (\bibinfo {year} {2016})}\BibitemShut
  {NoStop}%
\bibitem [{\citenamefont {Martinez}\ \emph {et~al.}(2016)\citenamefont
  {Martinez}, \citenamefont {Muschik}, \citenamefont {Schindler}, \citenamefont
  {Nigg}, \citenamefont {Erhard}, \citenamefont {Heyl}, \citenamefont {Hauke},
  \citenamefont {Dalmonte}, \citenamefont {Monz}, \citenamefont {Zoller} \emph
  {et~al.}}]{Martinez2016}%
  \BibitemOpen
  \bibfield  {author} {\bibinfo {author} {\bibfnamefont {E.~A.}\ \bibnamefont
  {Martinez}}, \bibinfo {author} {\bibfnamefont {C.~A.}\ \bibnamefont
  {Muschik}}, \bibinfo {author} {\bibfnamefont {P.}~\bibnamefont {Schindler}},
  \bibinfo {author} {\bibfnamefont {D.}~\bibnamefont {Nigg}}, \bibinfo {author}
  {\bibfnamefont {A.}~\bibnamefont {Erhard}}, \bibinfo {author} {\bibfnamefont
  {M.}~\bibnamefont {Heyl}}, \bibinfo {author} {\bibfnamefont {P.}~\bibnamefont
  {Hauke}}, \bibinfo {author} {\bibfnamefont {M.}~\bibnamefont {Dalmonte}},
  \bibinfo {author} {\bibfnamefont {T.}~\bibnamefont {Monz}}, \bibinfo {author}
  {\bibfnamefont {P.}~\bibnamefont {Zoller}},  \emph {et~al.},\ }\href
  {https://www.nature.com/articles/nature18318} {\bibfield  {journal} {\bibinfo
   {journal} {Nature}\ }\textbf {\bibinfo {volume} {534}},\ \bibinfo {pages}
  {516} (\bibinfo {year} {2016})}\BibitemShut {NoStop}%
\bibitem [{\citenamefont {Pichler}\ \emph {et~al.}(2016)\citenamefont
  {Pichler}, \citenamefont {Dalmonte}, \citenamefont {Rico}, \citenamefont
  {Zoller},\ and\ \citenamefont {Montangero}}]{Pichler2016}%
  \BibitemOpen
  \bibfield  {author} {\bibinfo {author} {\bibfnamefont {T.}~\bibnamefont
  {Pichler}}, \bibinfo {author} {\bibfnamefont {M.}~\bibnamefont {Dalmonte}},
  \bibinfo {author} {\bibfnamefont {E.}~\bibnamefont {Rico}}, \bibinfo {author}
  {\bibfnamefont {P.}~\bibnamefont {Zoller}}, \ and\ \bibinfo {author}
  {\bibfnamefont {S.}~\bibnamefont {Montangero}},\ }\href {\doibase
  10.1103/PhysRevX.6.011023} {\bibfield  {journal} {\bibinfo  {journal} {Phys.
  Rev. X}\ }\textbf {\bibinfo {volume} {6}},\ \bibinfo {pages} {011023}
  (\bibinfo {year} {2016})}\BibitemShut {NoStop}%
\bibitem [{\citenamefont {Schwinger}(1951)}]{Schwinger1951}%
  \BibitemOpen
  \bibfield  {author} {\bibinfo {author} {\bibfnamefont {J.}~\bibnamefont
  {Schwinger}},\ }\href {\doibase 10.1103/PhysRev.82.664} {\bibfield  {journal}
  {\bibinfo  {journal} {Phys. Rev.}\ }\textbf {\bibinfo {volume} {82}},\
  \bibinfo {pages} {664} (\bibinfo {year} {1951})}\BibitemShut {NoStop}%
\bibitem [{\citenamefont {Coleman}\ \emph {et~al.}(1975)\citenamefont
  {Coleman}, \citenamefont {Jackiw},\ and\ \citenamefont
  {Susskind}}]{Coleman1975charge}%
  \BibitemOpen
  \bibfield  {author} {\bibinfo {author} {\bibfnamefont {S.}~\bibnamefont
  {Coleman}}, \bibinfo {author} {\bibfnamefont {R.}~\bibnamefont {Jackiw}}, \
  and\ \bibinfo {author} {\bibfnamefont {L.}~\bibnamefont {Susskind}},\ }\href
  {https://www.sciencedirect.com/science/article/pii/0003491675902122}
  {\bibfield  {journal} {\bibinfo  {journal} {Annals of Physics}\ }\textbf
  {\bibinfo {volume} {93}},\ \bibinfo {pages} {267} (\bibinfo {year}
  {1975})}\BibitemShut {NoStop}%
\bibitem [{\citenamefont {Coleman}(1976)}]{Coleman1976more}%
  \BibitemOpen
  \bibfield  {author} {\bibinfo {author} {\bibfnamefont {S.}~\bibnamefont
  {Coleman}},\ }\href
  {https://www.sciencedirect.com/science/article/pii/0003491676902803}
  {\bibfield  {journal} {\bibinfo  {journal} {Annals of Physics}\ }\textbf
  {\bibinfo {volume} {101}},\ \bibinfo {pages} {239} (\bibinfo {year}
  {1976})}\BibitemShut {NoStop}%
\bibitem [{\citenamefont {Kogut}\ and\ \citenamefont
  {Stephanov}(2003)}]{Kogut2003}%
  \BibitemOpen
  \bibfield  {author} {\bibinfo {author} {\bibfnamefont {J.~B.}\ \bibnamefont
  {Kogut}}\ and\ \bibinfo {author} {\bibfnamefont {M.~A.}\ \bibnamefont
  {Stephanov}},\ }\href@noop {} {\emph {\bibinfo {title} {The phases of quantum
  chromodynamics: from confinement to extreme environments}}},\ Vol.~\bibinfo
  {volume} {21}\ (\bibinfo  {publisher} {Cambridge University Press},\ \bibinfo
  {year} {2003})\BibitemShut {NoStop}%
\bibitem [{\citenamefont {Heyl}\ \emph {et~al.}(2013)\citenamefont {Heyl},
  \citenamefont {Polkovnikov},\ and\ \citenamefont {Kehrein}}]{Markus2013}%
  \BibitemOpen
  \bibfield  {author} {\bibinfo {author} {\bibfnamefont {M.}~\bibnamefont
  {Heyl}}, \bibinfo {author} {\bibfnamefont {A.}~\bibnamefont {Polkovnikov}}, \
  and\ \bibinfo {author} {\bibfnamefont {S.}~\bibnamefont {Kehrein}},\ }\href
  {\doibase 10.1103/PhysRevLett.110.135704} {\bibfield  {journal} {\bibinfo
  {journal} {Phys. Rev. Lett.}\ }\textbf {\bibinfo {volume} {110}},\ \bibinfo
  {pages} {135704} (\bibinfo {year} {2013})}\BibitemShut {NoStop}%
\bibitem [{\citenamefont {Heyl}(2018)}]{Heyl2018Rev}%
  \BibitemOpen
  \bibfield  {author} {\bibinfo {author} {\bibfnamefont {M.}~\bibnamefont
  {Heyl}},\ }\href {http://iopscience.iop.org/article/10.1088/1361-6633/aaaf9a}
  {\bibfield  {journal} {\bibinfo  {journal} {Reports on Progress in Physics}\
  }\textbf {\bibinfo {volume} {81}},\ \bibinfo {pages} {054001} (\bibinfo
  {year} {2018})}\BibitemShut {NoStop}%
\bibitem [{\citenamefont {Jurcevic}\ \emph {et~al.}(2017)\citenamefont
  {Jurcevic}, \citenamefont {Shen}, \citenamefont {Hauke}, \citenamefont
  {Maier}, \citenamefont {Brydges}, \citenamefont {Hempel}, \citenamefont
  {Lanyon}, \citenamefont {Heyl}, \citenamefont {Blatt},\ and\ \citenamefont
  {Roos}}]{Jurcevic2017}%
  \BibitemOpen
  \bibfield  {author} {\bibinfo {author} {\bibfnamefont {P.}~\bibnamefont
  {Jurcevic}}, \bibinfo {author} {\bibfnamefont {H.}~\bibnamefont {Shen}},
  \bibinfo {author} {\bibfnamefont {P.}~\bibnamefont {Hauke}}, \bibinfo
  {author} {\bibfnamefont {C.}~\bibnamefont {Maier}}, \bibinfo {author}
  {\bibfnamefont {T.}~\bibnamefont {Brydges}}, \bibinfo {author} {\bibfnamefont
  {C.}~\bibnamefont {Hempel}}, \bibinfo {author} {\bibfnamefont {B.~P.}\
  \bibnamefont {Lanyon}}, \bibinfo {author} {\bibfnamefont {M.}~\bibnamefont
  {Heyl}}, \bibinfo {author} {\bibfnamefont {R.}~\bibnamefont {Blatt}}, \ and\
  \bibinfo {author} {\bibfnamefont {C.~F.}\ \bibnamefont {Roos}},\ }\href
  {\doibase 10.1103/PhysRevLett.119.080501} {\bibfield  {journal} {\bibinfo
  {journal} {Phys. Rev. Lett.}\ }\textbf {\bibinfo {volume} {119}},\ \bibinfo
  {pages} {080501} (\bibinfo {year} {2017})}\BibitemShut {NoStop}%
\bibitem [{\citenamefont {Zhang}\ \emph {et~al.}(2017)\citenamefont {Zhang},
  \citenamefont {Pagano}, \citenamefont {Hess}, \citenamefont {Kyprianidis},
  \citenamefont {Becker}, \citenamefont {Kaplan}, \citenamefont {Gorshkov},
  \citenamefont {Gong},\ and\ \citenamefont {Monroe}}]{Zhang2017}%
  \BibitemOpen
  \bibfield  {author} {\bibinfo {author} {\bibfnamefont {J.}~\bibnamefont
  {Zhang}}, \bibinfo {author} {\bibfnamefont {G.}~\bibnamefont {Pagano}},
  \bibinfo {author} {\bibfnamefont {P.~W.}\ \bibnamefont {Hess}}, \bibinfo
  {author} {\bibfnamefont {A.}~\bibnamefont {Kyprianidis}}, \bibinfo {author}
  {\bibfnamefont {P.}~\bibnamefont {Becker}}, \bibinfo {author} {\bibfnamefont
  {H.}~\bibnamefont {Kaplan}}, \bibinfo {author} {\bibfnamefont {A.~V.}\
  \bibnamefont {Gorshkov}}, \bibinfo {author} {\bibfnamefont {Z.-X.}\
  \bibnamefont {Gong}}, \ and\ \bibinfo {author} {\bibfnamefont
  {C.}~\bibnamefont {Monroe}},\ }\href
  {https://www.nature.com/articles/nature24654} {\bibfield  {journal} {\bibinfo
   {journal} {Nature}\ }\textbf {\bibinfo {volume} {551}},\ \bibinfo {pages}
  {601} (\bibinfo {year} {2017})}\BibitemShut {NoStop}%
\bibitem [{\citenamefont {Chandrasekharan}\ and\ \citenamefont
  {Wiese}(1997)}]{Chandrasekharan1997}%
  \BibitemOpen
  \bibfield  {author} {\bibinfo {author} {\bibfnamefont {S.}~\bibnamefont
  {Chandrasekharan}}\ and\ \bibinfo {author} {\bibfnamefont {U.-J.}\
  \bibnamefont {Wiese}},\ }\href {\doibase
  http://dx.doi.org/10.1016/S0550-3213(97)80041-7} {\bibfield  {journal}
  {\bibinfo  {journal} {Nuclear Physics B}\ }\textbf {\bibinfo {volume}
  {492}},\ \bibinfo {pages} {455 } (\bibinfo {year} {1997})}\BibitemShut
  {NoStop}%
\bibitem [{\citenamefont {Moessner}\ \emph {et~al.}(2001)\citenamefont
  {Moessner}, \citenamefont {Sondhi},\ and\ \citenamefont
  {Fradkin}}]{Moessner2001a}%
  \BibitemOpen
  \bibfield  {author} {\bibinfo {author} {\bibfnamefont {R.}~\bibnamefont
  {Moessner}}, \bibinfo {author} {\bibfnamefont {S.~L.}\ \bibnamefont
  {Sondhi}}, \ and\ \bibinfo {author} {\bibfnamefont {E.}~\bibnamefont
  {Fradkin}},\ }\href {\doibase 10.1103/PhysRevB.65.024504} {\bibfield
  {journal} {\bibinfo  {journal} {Phys. Rev. B}\ }\textbf {\bibinfo {volume}
  {65}},\ \bibinfo {pages} {024504} (\bibinfo {year} {2001})}\BibitemShut
  {NoStop}%
\bibitem [{\citenamefont {Moessner}\ and\ \citenamefont
  {Sondhi}(2001)}]{Moessner2001b}%
  \BibitemOpen
  \bibfield  {author} {\bibinfo {author} {\bibfnamefont {R.}~\bibnamefont
  {Moessner}}\ and\ \bibinfo {author} {\bibfnamefont {S.~L.}\ \bibnamefont
  {Sondhi}},\ }\href {\doibase 10.1103/PhysRevLett.86.1881} {\bibfield
  {journal} {\bibinfo  {journal} {Phys. Rev. Lett.}\ }\textbf {\bibinfo
  {volume} {86}},\ \bibinfo {pages} {1881} (\bibinfo {year}
  {2001})}\BibitemShut {NoStop}%
\bibitem [{\citenamefont {Huse}\ \emph {et~al.}(2003)\citenamefont {Huse},
  \citenamefont {Krauth}, \citenamefont {Moessner},\ and\ \citenamefont
  {Sondhi}}]{Huse2003}%
  \BibitemOpen
  \bibfield  {author} {\bibinfo {author} {\bibfnamefont {D.~A.}\ \bibnamefont
  {Huse}}, \bibinfo {author} {\bibfnamefont {W.}~\bibnamefont {Krauth}},
  \bibinfo {author} {\bibfnamefont {R.}~\bibnamefont {Moessner}}, \ and\
  \bibinfo {author} {\bibfnamefont {S.~L.}\ \bibnamefont {Sondhi}},\ }\href
  {\doibase 10.1103/PhysRevLett.91.167004} {\bibfield  {journal} {\bibinfo
  {journal} {Phys. Rev. Lett.}\ }\textbf {\bibinfo {volume} {91}},\ \bibinfo
  {pages} {167004} (\bibinfo {year} {2003})}\BibitemShut {NoStop}%
\bibitem [{\citenamefont {Shannon}\ \emph {et~al.}(2004)\citenamefont
  {Shannon}, \citenamefont {Misguich},\ and\ \citenamefont
  {Penc}}]{Shannon2004}%
  \BibitemOpen
  \bibfield  {author} {\bibinfo {author} {\bibfnamefont {N.}~\bibnamefont
  {Shannon}}, \bibinfo {author} {\bibfnamefont {G.}~\bibnamefont {Misguich}}, \
  and\ \bibinfo {author} {\bibfnamefont {K.}~\bibnamefont {Penc}},\ }\href
  {\doibase 10.1103/PhysRevB.69.220403} {\bibfield  {journal} {\bibinfo
  {journal} {Phys. Rev. B}\ }\textbf {\bibinfo {volume} {69}},\ \bibinfo
  {pages} {220403} (\bibinfo {year} {2004})}\BibitemShut {NoStop}%
\bibitem [{\citenamefont {Chakravarty}(2002)}]{Chakravarty2002}%
  \BibitemOpen
  \bibfield  {author} {\bibinfo {author} {\bibfnamefont {S.}~\bibnamefont
  {Chakravarty}},\ }\href {\doibase 10.1103/PhysRevB.66.224505} {\bibfield
  {journal} {\bibinfo  {journal} {Phys. Rev. B}\ }\textbf {\bibinfo {volume}
  {66}},\ \bibinfo {pages} {224505} (\bibinfo {year} {2002})}\BibitemShut
  {NoStop}%
\bibitem [{\citenamefont {Banerjee}\ \emph
  {et~al.}(2013{\natexlab{b}})\citenamefont {Banerjee}, \citenamefont {Jiang},
  \citenamefont {Widmer},\ and\ \citenamefont {Wiese}}]{Banerjee2013b}%
  \BibitemOpen
  \bibfield  {author} {\bibinfo {author} {\bibfnamefont {D.}~\bibnamefont
  {Banerjee}}, \bibinfo {author} {\bibfnamefont {F.-J.}\ \bibnamefont {Jiang}},
  \bibinfo {author} {\bibfnamefont {P.}~\bibnamefont {Widmer}}, \ and\ \bibinfo
  {author} {\bibfnamefont {U.-J.}\ \bibnamefont {Wiese}},\ }\href
  {http://stacks.iop.org/1742-5468/2013/i=12/a=P12010} {\bibfield  {journal}
  {\bibinfo  {journal} {Journal of Statistical Mechanics: Theory and
  Experiment}\ }\textbf {\bibinfo {volume} {2013}},\ \bibinfo {pages} {P12010}
  (\bibinfo {year} {2013}{\natexlab{b}})}\BibitemShut {NoStop}%
\bibitem [{\citenamefont {Banerjee}\ \emph {et~al.}(2018)\citenamefont
  {Banerjee}, \citenamefont {Jiang}, \citenamefont {Olesen}, \citenamefont
  {Orland},\ and\ \citenamefont {Wiese}}]{Banerjee2018}%
  \BibitemOpen
  \bibfield  {author} {\bibinfo {author} {\bibfnamefont {D.}~\bibnamefont
  {Banerjee}}, \bibinfo {author} {\bibfnamefont {F.-J.}\ \bibnamefont {Jiang}},
  \bibinfo {author} {\bibfnamefont {T.~Z.}\ \bibnamefont {Olesen}}, \bibinfo
  {author} {\bibfnamefont {P.}~\bibnamefont {Orland}}, \ and\ \bibinfo {author}
  {\bibfnamefont {U.-J.}\ \bibnamefont {Wiese}},\ }\href {\doibase
  10.1103/PhysRevB.97.205108} {\bibfield  {journal} {\bibinfo  {journal} {Phys.
  Rev. B}\ }\textbf {\bibinfo {volume} {97}},\ \bibinfo {pages} {205108}
  (\bibinfo {year} {2018})}\BibitemShut {NoStop}%
\bibitem [{\citenamefont {Banerjee}\ \emph {et~al.}(2014)\citenamefont
  {Banerjee}, \citenamefont {B\"ogli}, \citenamefont {Hofmann}, \citenamefont
  {Jiang}, \citenamefont {Widmer},\ and\ \citenamefont {Wiese}}]{Banerjee2014}%
  \BibitemOpen
  \bibfield  {author} {\bibinfo {author} {\bibfnamefont {D.}~\bibnamefont
  {Banerjee}}, \bibinfo {author} {\bibfnamefont {M.}~\bibnamefont {B\"ogli}},
  \bibinfo {author} {\bibfnamefont {C.~P.}\ \bibnamefont {Hofmann}}, \bibinfo
  {author} {\bibfnamefont {F.-J.}\ \bibnamefont {Jiang}}, \bibinfo {author}
  {\bibfnamefont {P.}~\bibnamefont {Widmer}}, \ and\ \bibinfo {author}
  {\bibfnamefont {U.-J.}\ \bibnamefont {Wiese}},\ }\href {\doibase
  10.1103/PhysRevB.90.245143} {\bibfield  {journal} {\bibinfo  {journal} {Phys.
  Rev. B}\ }\textbf {\bibinfo {volume} {90}},\ \bibinfo {pages} {245143}
  (\bibinfo {year} {2014})}\BibitemShut {NoStop}%
\bibitem [{\citenamefont {Rokhsar}\ and\ \citenamefont
  {Kivelson}(1988)}]{Rokhsar1988}%
  \BibitemOpen
  \bibfield  {author} {\bibinfo {author} {\bibfnamefont {D.~S.}\ \bibnamefont
  {Rokhsar}}\ and\ \bibinfo {author} {\bibfnamefont {S.~A.}\ \bibnamefont
  {Kivelson}},\ }\href {\doibase 10.1103/PhysRevLett.61.2376} {\bibfield
  {journal} {\bibinfo  {journal} {Phys. Rev. Lett.}\ }\textbf {\bibinfo
  {volume} {61}},\ \bibinfo {pages} {2376} (\bibinfo {year}
  {1988})}\BibitemShut {NoStop}%
\bibitem [{\citenamefont {Brower}\ \emph {et~al.}(1999)\citenamefont {Brower},
  \citenamefont {Chandrasekharan},\ and\ \citenamefont {Wiese}}]{Brower1999}%
  \BibitemOpen
  \bibfield  {author} {\bibinfo {author} {\bibfnamefont {R.}~\bibnamefont
  {Brower}}, \bibinfo {author} {\bibfnamefont {S.}~\bibnamefont
  {Chandrasekharan}}, \ and\ \bibinfo {author} {\bibfnamefont {U.-J.}\
  \bibnamefont {Wiese}},\ }\href {\doibase 10.1103/PhysRevD.60.094502}
  {\bibfield  {journal} {\bibinfo  {journal} {Phys. Rev. D}\ }\textbf {\bibinfo
  {volume} {60}},\ \bibinfo {pages} {094502} (\bibinfo {year}
  {1999})}\BibitemShut {NoStop}%
\bibitem [{\citenamefont {Lee}\ and\ \citenamefont {Yang}(1952)}]{Lee1952}%
  \BibitemOpen
  \bibfield  {author} {\bibinfo {author} {\bibfnamefont {T.~D.}\ \bibnamefont
  {Lee}}\ and\ \bibinfo {author} {\bibfnamefont {C.~N.}\ \bibnamefont {Yang}},\
  }\href {\doibase 10.1103/PhysRev.87.410} {\bibfield  {journal} {\bibinfo
  {journal} {Phys. Rev.}\ }\textbf {\bibinfo {volume} {87}},\ \bibinfo {pages}
  {410} (\bibinfo {year} {1952})}\BibitemShut {NoStop}%
\bibitem [{\citenamefont {Yang}\ and\ \citenamefont {Lee}(1952)}]{Yang1952}%
  \BibitemOpen
  \bibfield  {author} {\bibinfo {author} {\bibfnamefont {C.~N.}\ \bibnamefont
  {Yang}}\ and\ \bibinfo {author} {\bibfnamefont {T.~D.}\ \bibnamefont {Lee}},\
  }\href {\doibase 10.1103/PhysRev.87.404} {\bibfield  {journal} {\bibinfo
  {journal} {Phys. Rev.}\ }\textbf {\bibinfo {volume} {87}},\ \bibinfo {pages}
  {404} (\bibinfo {year} {1952})}\BibitemShut {NoStop}%
\bibitem [{\citenamefont {Fisher}(1965)}]{Fisher1965nature}%
  \BibitemOpen
  \bibfield  {author} {\bibinfo {author} {\bibfnamefont {M.}~\bibnamefont
  {Fisher}},\ }\href@noop {} {\emph {\bibinfo {title} {The Nature of Critical
  Points (Lectures in Theoretical Physics vol 7C) ed Brittin WE and Dunham
  LG}}}\ (\bibinfo  {publisher} {the University of Colorado Press, Boulder},\
  \bibinfo {year} {1965})\BibitemShut {NoStop}%
\bibitem [{\citenamefont {Karrasch}\ and\ \citenamefont
  {Schuricht}(2013)}]{Karrasch2013}%
  \BibitemOpen
  \bibfield  {author} {\bibinfo {author} {\bibfnamefont {C.}~\bibnamefont
  {Karrasch}}\ and\ \bibinfo {author} {\bibfnamefont {D.}~\bibnamefont
  {Schuricht}},\ }\href {\doibase 10.1103/PhysRevB.87.195104} {\bibfield
  {journal} {\bibinfo  {journal} {Phys. Rev. B}\ }\textbf {\bibinfo {volume}
  {87}},\ \bibinfo {pages} {195104} (\bibinfo {year} {2013})}\BibitemShut
  {NoStop}%
\bibitem [{\citenamefont {Kriel}\ \emph {et~al.}(2014)\citenamefont {Kriel},
  \citenamefont {Karrasch},\ and\ \citenamefont {Kehrein}}]{Kriel2014}%
  \BibitemOpen
  \bibfield  {author} {\bibinfo {author} {\bibfnamefont {J.~N.}\ \bibnamefont
  {Kriel}}, \bibinfo {author} {\bibfnamefont {C.}~\bibnamefont {Karrasch}}, \
  and\ \bibinfo {author} {\bibfnamefont {S.}~\bibnamefont {Kehrein}},\ }\href
  {\doibase 10.1103/PhysRevB.90.125106} {\bibfield  {journal} {\bibinfo
  {journal} {Phys. Rev. B}\ }\textbf {\bibinfo {volume} {90}},\ \bibinfo
  {pages} {125106} (\bibinfo {year} {2014})}\BibitemShut {NoStop}%
\bibitem [{\citenamefont {Heyl}(2015)}]{Markus2015}%
  \BibitemOpen
  \bibfield  {author} {\bibinfo {author} {\bibfnamefont {M.}~\bibnamefont
  {Heyl}},\ }\href {\doibase 10.1103/PhysRevLett.115.140602} {\bibfield
  {journal} {\bibinfo  {journal} {Phys. Rev. Lett.}\ }\textbf {\bibinfo
  {volume} {115}},\ \bibinfo {pages} {140602} (\bibinfo {year}
  {2015})}\BibitemShut {NoStop}%
\bibitem [{\citenamefont {Sharma}\ \emph {et~al.}(2015)\citenamefont {Sharma},
  \citenamefont {Suzuki},\ and\ \citenamefont {Dutta}}]{Sharma2015}%
  \BibitemOpen
  \bibfield  {author} {\bibinfo {author} {\bibfnamefont {S.}~\bibnamefont
  {Sharma}}, \bibinfo {author} {\bibfnamefont {S.}~\bibnamefont {Suzuki}}, \
  and\ \bibinfo {author} {\bibfnamefont {A.}~\bibnamefont {Dutta}},\ }\href
  {\doibase 10.1103/PhysRevB.92.104306} {\bibfield  {journal} {\bibinfo
  {journal} {Phys. Rev. B}\ }\textbf {\bibinfo {volume} {92}},\ \bibinfo
  {pages} {104306} (\bibinfo {year} {2015})}\BibitemShut {NoStop}%
\bibitem [{\citenamefont {Budich}\ and\ \citenamefont
  {Heyl}(2016)}]{Budich2016}%
  \BibitemOpen
  \bibfield  {author} {\bibinfo {author} {\bibfnamefont {J.~C.}\ \bibnamefont
  {Budich}}\ and\ \bibinfo {author} {\bibfnamefont {M.}~\bibnamefont {Heyl}},\
  }\href {\doibase 10.1103/PhysRevB.93.085416} {\bibfield  {journal} {\bibinfo
  {journal} {Phys. Rev. B}\ }\textbf {\bibinfo {volume} {93}},\ \bibinfo
  {pages} {085416} (\bibinfo {year} {2016})}\BibitemShut {NoStop}%
\bibitem [{\citenamefont {Bhattacharya}\ \emph {et~al.}(2017)\citenamefont
  {Bhattacharya}, \citenamefont {Bandyopadhyay},\ and\ \citenamefont
  {Dutta}}]{Bhattacharya2017}%
  \BibitemOpen
  \bibfield  {author} {\bibinfo {author} {\bibfnamefont {U.}~\bibnamefont
  {Bhattacharya}}, \bibinfo {author} {\bibfnamefont {S.}~\bibnamefont
  {Bandyopadhyay}}, \ and\ \bibinfo {author} {\bibfnamefont {A.}~\bibnamefont
  {Dutta}},\ }\href {\doibase 10.1103/PhysRevB.96.180303} {\bibfield  {journal}
  {\bibinfo  {journal} {Phys. Rev. B}\ }\textbf {\bibinfo {volume} {96}},\
  \bibinfo {pages} {180303} (\bibinfo {year} {2017})}\BibitemShut {NoStop}%
\bibitem [{\citenamefont {Fl{\"a}schner}\ \emph {et~al.}(2018)\citenamefont
  {Fl{\"a}schner}, \citenamefont {Vogel}, \citenamefont {Tarnowski},
  \citenamefont {Rem}, \citenamefont {L{\"u}hmann}, \citenamefont {Heyl},
  \citenamefont {Budich}, \citenamefont {Mathey}, \citenamefont {Sengstock},\
  and\ \citenamefont {Weitenberg}}]{Flaschner2018}%
  \BibitemOpen
  \bibfield  {author} {\bibinfo {author} {\bibfnamefont {N.}~\bibnamefont
  {Fl{\"a}schner}}, \bibinfo {author} {\bibfnamefont {D.}~\bibnamefont
  {Vogel}}, \bibinfo {author} {\bibfnamefont {M.}~\bibnamefont {Tarnowski}},
  \bibinfo {author} {\bibfnamefont {B.}~\bibnamefont {Rem}}, \bibinfo {author}
  {\bibfnamefont {D.-S.}\ \bibnamefont {L{\"u}hmann}}, \bibinfo {author}
  {\bibfnamefont {M.}~\bibnamefont {Heyl}}, \bibinfo {author} {\bibfnamefont
  {J.}~\bibnamefont {Budich}}, \bibinfo {author} {\bibfnamefont
  {L.}~\bibnamefont {Mathey}}, \bibinfo {author} {\bibfnamefont
  {K.}~\bibnamefont {Sengstock}}, \ and\ \bibinfo {author} {\bibfnamefont
  {C.}~\bibnamefont {Weitenberg}},\ }\href
  {https://www.nature.com/articles/s41567-017-0013-8} {\bibfield  {journal}
  {\bibinfo  {journal} {Nature Physics}\ }\textbf {\bibinfo {volume} {14}},\
  \bibinfo {pages} {265} (\bibinfo {year} {2018})}\BibitemShut {NoStop}%
\bibitem [{\citenamefont {Xu}\ \emph {et~al.}(2018)\citenamefont {Xu},
  \citenamefont {Wang}, \citenamefont {Heyl}, \citenamefont {Budich},
  \citenamefont {Pan}, \citenamefont {Chen}, \citenamefont {Jan}, \citenamefont
  {Sun}, \citenamefont {Xu}, \citenamefont {Han} \emph {et~al.}}]{xu2018}%
  \BibitemOpen
  \bibfield  {author} {\bibinfo {author} {\bibfnamefont {X.-Y.}\ \bibnamefont
  {Xu}}, \bibinfo {author} {\bibfnamefont {Q.-Q.}\ \bibnamefont {Wang}},
  \bibinfo {author} {\bibfnamefont {M.}~\bibnamefont {Heyl}}, \bibinfo {author}
  {\bibfnamefont {J.~C.}\ \bibnamefont {Budich}}, \bibinfo {author}
  {\bibfnamefont {W.-W.}\ \bibnamefont {Pan}}, \bibinfo {author} {\bibfnamefont
  {Z.}~\bibnamefont {Chen}}, \bibinfo {author} {\bibfnamefont {M.}~\bibnamefont
  {Jan}}, \bibinfo {author} {\bibfnamefont {K.}~\bibnamefont {Sun}}, \bibinfo
  {author} {\bibfnamefont {J.-S.}\ \bibnamefont {Xu}}, \bibinfo {author}
  {\bibfnamefont {Y.-J.}\ \bibnamefont {Han}},  \emph {et~al.},\ }\href
  {https://arxiv.org/abs/1808.03930} {\bibfield  {journal} {\bibinfo  {journal}
  {arXiv preprint arXiv:1808.03930}\ } (\bibinfo {year} {2018})}\BibitemShut
  {NoStop}%
\bibitem [{\citenamefont {Guo}\ \emph {et~al.}(2018)\citenamefont {Guo},
  \citenamefont {Yang}, \citenamefont {Zeng}, \citenamefont {Peng},
  \citenamefont {Li}, \citenamefont {Deng}, \citenamefont {Jin}, \citenamefont
  {Chen}, \citenamefont {Zheng},\ and\ \citenamefont {Fan}}]{guo2018}%
  \BibitemOpen
  \bibfield  {author} {\bibinfo {author} {\bibfnamefont {X.-Y.}\ \bibnamefont
  {Guo}}, \bibinfo {author} {\bibfnamefont {C.}~\bibnamefont {Yang}}, \bibinfo
  {author} {\bibfnamefont {Y.}~\bibnamefont {Zeng}}, \bibinfo {author}
  {\bibfnamefont {Y.}~\bibnamefont {Peng}}, \bibinfo {author} {\bibfnamefont
  {H.-K.}\ \bibnamefont {Li}}, \bibinfo {author} {\bibfnamefont
  {H.}~\bibnamefont {Deng}}, \bibinfo {author} {\bibfnamefont {Y.-R.}\
  \bibnamefont {Jin}}, \bibinfo {author} {\bibfnamefont {S.}~\bibnamefont
  {Chen}}, \bibinfo {author} {\bibfnamefont {D.}~\bibnamefont {Zheng}}, \ and\
  \bibinfo {author} {\bibfnamefont {H.}~\bibnamefont {Fan}},\ }\href
  {https://arxiv.org/abs/1806.09269} {\bibfield  {journal} {\bibinfo  {journal}
  {arXiv preprint arXiv:1806.09269}\ } (\bibinfo {year} {2018})}\BibitemShut
  {NoStop}%
\bibitem [{\citenamefont {Wang}\ \emph {et~al.}(2018)\citenamefont {Wang},
  \citenamefont {Qiu}, \citenamefont {Xiao}, \citenamefont {Zhan},
  \citenamefont {Bian}, \citenamefont {Yi},\ and\ \citenamefont
  {Xue}}]{wang2018}%
  \BibitemOpen
  \bibfield  {author} {\bibinfo {author} {\bibfnamefont {K.}~\bibnamefont
  {Wang}}, \bibinfo {author} {\bibfnamefont {X.}~\bibnamefont {Qiu}}, \bibinfo
  {author} {\bibfnamefont {L.}~\bibnamefont {Xiao}}, \bibinfo {author}
  {\bibfnamefont {X.}~\bibnamefont {Zhan}}, \bibinfo {author} {\bibfnamefont
  {Z.}~\bibnamefont {Bian}}, \bibinfo {author} {\bibfnamefont {W.}~\bibnamefont
  {Yi}}, \ and\ \bibinfo {author} {\bibfnamefont {P.}~\bibnamefont {Xue}},\
  }\href {https://arxiv.org/abs/1806.10871} {\bibfield  {journal} {\bibinfo
  {journal} {arXiv preprint arXiv:1806.10871}\ } (\bibinfo {year}
  {2018})}\BibitemShut {NoStop}%
\bibitem [{\citenamefont {Smale}\ \emph {et~al.}(2018)\citenamefont {Smale},
  \citenamefont {He}, \citenamefont {Olsen}, \citenamefont {Jackson},
  \citenamefont {Sharum}, \citenamefont {Trotzky}, \citenamefont {Marino},
  \citenamefont {Rey},\ and\ \citenamefont {Thywissen}}]{smale2018}%
  \BibitemOpen
  \bibfield  {author} {\bibinfo {author} {\bibfnamefont {S.}~\bibnamefont
  {Smale}}, \bibinfo {author} {\bibfnamefont {P.}~\bibnamefont {He}}, \bibinfo
  {author} {\bibfnamefont {B.~A.}\ \bibnamefont {Olsen}}, \bibinfo {author}
  {\bibfnamefont {K.~G.}\ \bibnamefont {Jackson}}, \bibinfo {author}
  {\bibfnamefont {H.}~\bibnamefont {Sharum}}, \bibinfo {author} {\bibfnamefont
  {S.}~\bibnamefont {Trotzky}}, \bibinfo {author} {\bibfnamefont
  {J.}~\bibnamefont {Marino}}, \bibinfo {author} {\bibfnamefont {A.~M.}\
  \bibnamefont {Rey}}, \ and\ \bibinfo {author} {\bibfnamefont {J.~H.}\
  \bibnamefont {Thywissen}},\ }\href {https://arxiv.org/abs/1806.11044}
  {\bibfield  {journal} {\bibinfo  {journal} {arXiv preprint arXiv:1806.11044}\
  } (\bibinfo {year} {2018})}\BibitemShut {NoStop}%
\bibitem [{\citenamefont {Tian}\ \emph {et~al.}(2018)\citenamefont {Tian},
  \citenamefont {Ke}, \citenamefont {Zhang}, \citenamefont {Lin}, \citenamefont
  {Shi}, \citenamefont {Huang}, \citenamefont {Lee},\ and\ \citenamefont
  {Du}}]{tian2018}%
  \BibitemOpen
  \bibfield  {author} {\bibinfo {author} {\bibfnamefont {T.}~\bibnamefont
  {Tian}}, \bibinfo {author} {\bibfnamefont {Y.}~\bibnamefont {Ke}}, \bibinfo
  {author} {\bibfnamefont {L.}~\bibnamefont {Zhang}}, \bibinfo {author}
  {\bibfnamefont {S.}~\bibnamefont {Lin}}, \bibinfo {author} {\bibfnamefont
  {Z.}~\bibnamefont {Shi}}, \bibinfo {author} {\bibfnamefont {P.}~\bibnamefont
  {Huang}}, \bibinfo {author} {\bibfnamefont {C.}~\bibnamefont {Lee}}, \ and\
  \bibinfo {author} {\bibfnamefont {J.}~\bibnamefont {Du}},\ }\href
  {https://arxiv.org/abs/1807.04483} {\bibfield  {journal} {\bibinfo  {journal}
  {arXiv preprint arXiv:1807.04483}\ } (\bibinfo {year} {2018})}\BibitemShut
  {NoStop}%
\bibitem [{\citenamefont {Heyl}(2014)}]{Heyl2014}%
  \BibitemOpen
  \bibfield  {author} {\bibinfo {author} {\bibfnamefont {M.}~\bibnamefont
  {Heyl}},\ }\href {\doibase 10.1103/PhysRevLett.113.205701} {\bibfield
  {journal} {\bibinfo  {journal} {Phys. Rev. Lett.}\ }\textbf {\bibinfo
  {volume} {113}},\ \bibinfo {pages} {205701} (\bibinfo {year}
  {2014})}\BibitemShut {NoStop}%
\bibitem [{\citenamefont {Weidinger}\ \emph {et~al.}(2017)\citenamefont
  {Weidinger}, \citenamefont {Heyl}, \citenamefont {Silva},\ and\ \citenamefont
  {Knap}}]{Weidinger2017}%
  \BibitemOpen
  \bibfield  {author} {\bibinfo {author} {\bibfnamefont {S.~A.}\ \bibnamefont
  {Weidinger}}, \bibinfo {author} {\bibfnamefont {M.}~\bibnamefont {Heyl}},
  \bibinfo {author} {\bibfnamefont {A.}~\bibnamefont {Silva}}, \ and\ \bibinfo
  {author} {\bibfnamefont {M.}~\bibnamefont {Knap}},\ }\href {\doibase
  10.1103/PhysRevB.96.134313} {\bibfield  {journal} {\bibinfo  {journal} {Phys.
  Rev. B}\ }\textbf {\bibinfo {volume} {96}},\ \bibinfo {pages} {134313}
  (\bibinfo {year} {2017})}\BibitemShut {NoStop}%
\bibitem [{\citenamefont {\ifmmode \check{Z}\else
  \v{Z}\fi{}unkovi\ifmmode~\check{c}\else \v{c}\fi{}}\ \emph
  {et~al.}(2018)\citenamefont {\ifmmode \check{Z}\else
  \v{Z}\fi{}unkovi\ifmmode~\check{c}\else \v{c}\fi{}}, \citenamefont {Heyl},
  \citenamefont {Knap},\ and\ \citenamefont {Silva}}]{Zunkovic2018}%
  \BibitemOpen
  \bibfield  {author} {\bibinfo {author} {\bibfnamefont {B.}~\bibnamefont
  {\ifmmode \check{Z}\else \v{Z}\fi{}unkovi\ifmmode~\check{c}\else
  \v{c}\fi{}}}, \bibinfo {author} {\bibfnamefont {M.}~\bibnamefont {Heyl}},
  \bibinfo {author} {\bibfnamefont {M.}~\bibnamefont {Knap}}, \ and\ \bibinfo
  {author} {\bibfnamefont {A.}~\bibnamefont {Silva}},\ }\href {\doibase
  10.1103/PhysRevLett.120.130601} {\bibfield  {journal} {\bibinfo  {journal}
  {Phys. Rev. Lett.}\ }\textbf {\bibinfo {volume} {120}},\ \bibinfo {pages}
  {130601} (\bibinfo {year} {2018})}\BibitemShut {NoStop}%
\bibitem [{\citenamefont {Vidal}(2003)}]{Vidal2003}%
  \BibitemOpen
  \bibfield  {author} {\bibinfo {author} {\bibfnamefont {G.}~\bibnamefont
  {Vidal}},\ }\href {\doibase 10.1103/PhysRevLett.91.147902} {\bibfield
  {journal} {\bibinfo  {journal} {Phys. Rev. Lett.}\ }\textbf {\bibinfo
  {volume} {91}},\ \bibinfo {pages} {147902} (\bibinfo {year}
  {2003})}\BibitemShut {NoStop}%
\bibitem [{\citenamefont {Schollw{\"o}ck}(2011)}]{Schollwock2011}%
  \BibitemOpen
  \bibfield  {author} {\bibinfo {author} {\bibfnamefont {U.}~\bibnamefont
  {Schollw{\"o}ck}},\ }\href
  {https://www.sciencedirect.com/science/article/pii/S0003491610001752?via%3Dihub}
  {\bibfield  {journal} {\bibinfo  {journal} {Annals of Physics}\ }\textbf
  {\bibinfo {volume} {326}},\ \bibinfo {pages} {96} (\bibinfo {year}
  {2011})}\BibitemShut {NoStop}%
\bibitem [{ite()}]{itensor}%
  \BibitemOpen
  \href@noop {} {}\bibinfo {note} {Calculations performed using the ITensor C++
  library, http://itensor.org/}\BibitemShut {NoStop}%
\bibitem [{\citenamefont {Sandvik}(2010)}]{sandvik2010}%
  \BibitemOpen
  \bibfield  {author} {\bibinfo {author} {\bibfnamefont {A.~W.}\ \bibnamefont
  {Sandvik}},\ }\bibfield  {booktitle} {\emph {\bibinfo {booktitle} {AIP
  Conference Proceedings}},\ }\href
  {https://aip.scitation.org/doi/10.1063/1.3518900} {\ \textbf {\bibinfo
  {volume} {1297}},\ \bibinfo {pages} {135} (\bibinfo {year}
  {2010})}\BibitemShut {NoStop}%
\bibitem [{\citenamefont {Byrnes}\ \emph {et~al.}(2002)\citenamefont {Byrnes},
  \citenamefont {Sriganesh}, \citenamefont {Bursill},\ and\ \citenamefont
  {Hamer}}]{Byrnes2002}%
  \BibitemOpen
  \bibfield  {author} {\bibinfo {author} {\bibfnamefont {T.~M.~R.}\
  \bibnamefont {Byrnes}}, \bibinfo {author} {\bibfnamefont {P.}~\bibnamefont
  {Sriganesh}}, \bibinfo {author} {\bibfnamefont {R.~J.}\ \bibnamefont
  {Bursill}}, \ and\ \bibinfo {author} {\bibfnamefont {C.~J.}\ \bibnamefont
  {Hamer}},\ }\href {\doibase 10.1103/PhysRevD.66.013002} {\bibfield  {journal}
  {\bibinfo  {journal} {Phys. Rev. D}\ }\textbf {\bibinfo {volume} {66}},\
  \bibinfo {pages} {013002} (\bibinfo {year} {2002})}\BibitemShut {NoStop}%
\bibitem [{Sup()}]{SupMat}%
  \BibitemOpen
  \href@noop {} {}\bibinfo {note} {See supplement materials.}\BibitemShut
  {Stop}%
\bibitem [{\citenamefont {Braguta}\ \emph {et~al.}(2012)\citenamefont
  {Braguta}, \citenamefont {Buividovich}, \citenamefont {Chernodub},
  \citenamefont {Kotov},\ and\ \citenamefont {Polikarpov}}]{braguta2012}%
  \BibitemOpen
  \bibfield  {author} {\bibinfo {author} {\bibfnamefont {V.}~\bibnamefont
  {Braguta}}, \bibinfo {author} {\bibfnamefont {P.}~\bibnamefont
  {Buividovich}}, \bibinfo {author} {\bibfnamefont {M.}~\bibnamefont
  {Chernodub}}, \bibinfo {author} {\bibfnamefont {A.~Y.}\ \bibnamefont
  {Kotov}}, \ and\ \bibinfo {author} {\bibfnamefont {M.}~\bibnamefont
  {Polikarpov}},\ }\href
  {https://www.sciencedirect.com/science/article/pii/S0370269312011471}
  {\bibfield  {journal} {\bibinfo  {journal} {Physics Letters B}\ }\textbf
  {\bibinfo {volume} {718}},\ \bibinfo {pages} {667} (\bibinfo {year}
  {2012})}\BibitemShut {NoStop}%
\bibitem [{\citenamefont {Bali}\ \emph {et~al.}(2012)\citenamefont {Bali},
  \citenamefont {Bruckmann}, \citenamefont {Endr\ifmmode~\mbox{\H{o}}\else
  \H{o}\fi{}di}, \citenamefont {Fodor}, \citenamefont {Katz},\ and\
  \citenamefont {Sch\"afer}}]{Bali2012}%
  \BibitemOpen
  \bibfield  {author} {\bibinfo {author} {\bibfnamefont {G.~S.}\ \bibnamefont
  {Bali}}, \bibinfo {author} {\bibfnamefont {F.}~\bibnamefont {Bruckmann}},
  \bibinfo {author} {\bibfnamefont {G.}~\bibnamefont
  {Endr\ifmmode~\mbox{\H{o}}\else \H{o}\fi{}di}}, \bibinfo {author}
  {\bibfnamefont {Z.}~\bibnamefont {Fodor}}, \bibinfo {author} {\bibfnamefont
  {S.~D.}\ \bibnamefont {Katz}}, \ and\ \bibinfo {author} {\bibfnamefont
  {A.}~\bibnamefont {Sch\"afer}},\ }\href {\doibase 10.1103/PhysRevD.86.071502}
  {\bibfield  {journal} {\bibinfo  {journal} {Phys. Rev. D}\ }\textbf {\bibinfo
  {volume} {86}},\ \bibinfo {pages} {071502} (\bibinfo {year}
  {2012})}\BibitemShut {NoStop}%
\bibitem [{\citenamefont {Zohar}\ \emph {et~al.}(2015)\citenamefont {Zohar},
  \citenamefont {Cirac},\ and\ \citenamefont {Reznik}}]{zohar2015}%
  \BibitemOpen
  \bibfield  {author} {\bibinfo {author} {\bibfnamefont {E.}~\bibnamefont
  {Zohar}}, \bibinfo {author} {\bibfnamefont {J.~I.}\ \bibnamefont {Cirac}}, \
  and\ \bibinfo {author} {\bibfnamefont {B.}~\bibnamefont {Reznik}},\ }\href
  {http://iopscience.iop.org/article/10.1088/0034-4885/79/1/014401/meta}
  {\bibfield  {journal} {\bibinfo  {journal} {Reports on Progress in Physics}\
  }\textbf {\bibinfo {volume} {79}},\ \bibinfo {pages} {014401} (\bibinfo
  {year} {2015})}\BibitemShut {NoStop}%
\bibitem [{\citenamefont {Marcos}\ \emph {et~al.}(2014)\citenamefont {Marcos},
  \citenamefont {Widmer}, \citenamefont {Rico}, \citenamefont {Hafezi},
  \citenamefont {Rabl}, \citenamefont {Wiese},\ and\ \citenamefont
  {Zoller}}]{marcos2014}%
  \BibitemOpen
  \bibfield  {author} {\bibinfo {author} {\bibfnamefont {D.}~\bibnamefont
  {Marcos}}, \bibinfo {author} {\bibfnamefont {P.}~\bibnamefont {Widmer}},
  \bibinfo {author} {\bibfnamefont {E.}~\bibnamefont {Rico}}, \bibinfo {author}
  {\bibfnamefont {M.}~\bibnamefont {Hafezi}}, \bibinfo {author} {\bibfnamefont
  {P.}~\bibnamefont {Rabl}}, \bibinfo {author} {\bibfnamefont {U.-J.}\
  \bibnamefont {Wiese}}, \ and\ \bibinfo {author} {\bibfnamefont
  {P.}~\bibnamefont {Zoller}},\ }\href
  {https://www.sciencedirect.com/science/article/pii/S0003491614002711?via%3Dihub}
  {\bibfield  {journal} {\bibinfo  {journal} {Annals of physics}\ }\textbf
  {\bibinfo {volume} {351}},\ \bibinfo {pages} {634} (\bibinfo {year}
  {2014})}\BibitemShut {NoStop}%
\bibitem [{\citenamefont {Zohar}\ and\ \citenamefont
  {Reznik}(2011)}]{Zohar2011}%
  \BibitemOpen
  \bibfield  {author} {\bibinfo {author} {\bibfnamefont {E.}~\bibnamefont
  {Zohar}}\ and\ \bibinfo {author} {\bibfnamefont {B.}~\bibnamefont {Reznik}},\
  }\href {\doibase 10.1103/PhysRevLett.107.275301} {\bibfield  {journal}
  {\bibinfo  {journal} {Phys. Rev. Lett.}\ }\textbf {\bibinfo {volume} {107}},\
  \bibinfo {pages} {275301} (\bibinfo {year} {2011})}\BibitemShut {NoStop}%
\bibitem [{\citenamefont {Tagliacozzo}\ \emph {et~al.}(2013)\citenamefont
  {Tagliacozzo}, \citenamefont {Celi}, \citenamefont {Zamora},\ and\
  \citenamefont {Lewenstein}}]{tagliacozzo2013}%
  \BibitemOpen
  \bibfield  {author} {\bibinfo {author} {\bibfnamefont {L.}~\bibnamefont
  {Tagliacozzo}}, \bibinfo {author} {\bibfnamefont {A.}~\bibnamefont {Celi}},
  \bibinfo {author} {\bibfnamefont {A.}~\bibnamefont {Zamora}}, \ and\ \bibinfo
  {author} {\bibfnamefont {M.}~\bibnamefont {Lewenstein}},\ }\href
  {https://www.sciencedirect.com/science/article/pii/S0003491612001819?via%3Dihub}
  {\bibfield  {journal} {\bibinfo  {journal} {Annals of Physics}\ }\textbf
  {\bibinfo {volume} {330}},\ \bibinfo {pages} {160} (\bibinfo {year}
  {2013})}\BibitemShut {NoStop}%
\bibitem [{\citenamefont {Lanyon}\ \emph {et~al.}(2011)\citenamefont {Lanyon},
  \citenamefont {Hempel}, \citenamefont {Nigg}, \citenamefont {M{\"u}ller},
  \citenamefont {Gerritsma}, \citenamefont {Z{\"a}hringer}, \citenamefont
  {Schindler}, \citenamefont {Barreiro}, \citenamefont {Rambach}, \citenamefont
  {Kirchmair} \emph {et~al.}}]{lanyon2011}%
  \BibitemOpen
  \bibfield  {author} {\bibinfo {author} {\bibfnamefont {B.~P.}\ \bibnamefont
  {Lanyon}}, \bibinfo {author} {\bibfnamefont {C.}~\bibnamefont {Hempel}},
  \bibinfo {author} {\bibfnamefont {D.}~\bibnamefont {Nigg}}, \bibinfo {author}
  {\bibfnamefont {M.}~\bibnamefont {M{\"u}ller}}, \bibinfo {author}
  {\bibfnamefont {R.}~\bibnamefont {Gerritsma}}, \bibinfo {author}
  {\bibfnamefont {F.}~\bibnamefont {Z{\"a}hringer}}, \bibinfo {author}
  {\bibfnamefont {P.}~\bibnamefont {Schindler}}, \bibinfo {author}
  {\bibfnamefont {J.}~\bibnamefont {Barreiro}}, \bibinfo {author}
  {\bibfnamefont {M.}~\bibnamefont {Rambach}}, \bibinfo {author} {\bibfnamefont
  {G.}~\bibnamefont {Kirchmair}},  \emph {et~al.},\ }\href
  {http://science.sciencemag.org/content/334/6052/57} {\bibfield  {journal}
  {\bibinfo  {journal} {Science}\ }\textbf {\bibinfo {volume} {334}},\ \bibinfo
  {pages} {57} (\bibinfo {year} {2011})}\BibitemShut {NoStop}%
\bibitem [{\citenamefont {Barends}\ \emph {et~al.}(2016)\citenamefont
  {Barends}, \citenamefont {Shabani}, \citenamefont {Lamata}, \citenamefont
  {Kelly}, \citenamefont {Mezzacapo}, \citenamefont {Las~Heras}, \citenamefont
  {Babbush}, \citenamefont {Fowler}, \citenamefont {Campbell}, \citenamefont
  {Chen} \emph {et~al.}}]{barends2016}%
  \BibitemOpen
  \bibfield  {author} {\bibinfo {author} {\bibfnamefont {R.}~\bibnamefont
  {Barends}}, \bibinfo {author} {\bibfnamefont {A.}~\bibnamefont {Shabani}},
  \bibinfo {author} {\bibfnamefont {L.}~\bibnamefont {Lamata}}, \bibinfo
  {author} {\bibfnamefont {J.}~\bibnamefont {Kelly}}, \bibinfo {author}
  {\bibfnamefont {A.}~\bibnamefont {Mezzacapo}}, \bibinfo {author}
  {\bibfnamefont {U.}~\bibnamefont {Las~Heras}}, \bibinfo {author}
  {\bibfnamefont {R.}~\bibnamefont {Babbush}}, \bibinfo {author} {\bibfnamefont
  {A.~G.}\ \bibnamefont {Fowler}}, \bibinfo {author} {\bibfnamefont
  {B.}~\bibnamefont {Campbell}}, \bibinfo {author} {\bibfnamefont
  {Y.}~\bibnamefont {Chen}},  \emph {et~al.},\ }\href
  {http://www.nature.com/articles/nature17658} {\bibfield  {journal} {\bibinfo
  {journal} {Nature}\ }\textbf {\bibinfo {volume} {534}},\ \bibinfo {pages}
  {222} (\bibinfo {year} {2016})}\BibitemShut {NoStop}%
\bibitem [{\citenamefont {Park}\ and\ \citenamefont
  {Light}(1986)}]{park1986unitary}%
  \BibitemOpen
  \bibfield  {author} {\bibinfo {author} {\bibfnamefont {T.~J.}\ \bibnamefont
  {Park}}\ and\ \bibinfo {author} {\bibfnamefont {J.}~\bibnamefont {Light}},\
  }\href {https://aip.scitation.org/doi/10.1063/1.451548} {\bibfield  {journal}
  {\bibinfo  {journal} {The Journal of chemical physics}\ }\textbf {\bibinfo
  {volume} {85}},\ \bibinfo {pages} {5870} (\bibinfo {year}
  {1986})}\BibitemShut {NoStop}%
\end{thebibliography}%

\clearpage
\newpage
\appendix

\section{TEBD}
\label{app:1d_qlm}
After the Jordan-Wigner transformation, the one-dimensional model is simplified as:
\begin{eqnarray}
	H^{1D}&=& \sum_{x}mp_x\left(\frac{1}{2}+\sigma^z_{x} \right)
	-\kappa\left(\sigma_{x}^{+}S_{x,\hat{i}}^{+}\sigma_{x+1}^{-} +h.c.\right)\text{.}\nonumber\\
	\label{H_1+1_d}
\end{eqnarray}
Here, $\sigma$ and $S$ are spin-$1/2$ operators, representing the matter fields on the sites,
and the gauge fields on the bonds, respectively.

\emph{Symmetries--}
The Hamiltonian is invariant under the symmetry $\mathcal{C}$ and $\mathcal{P}$.

\emph{Charge conjugation, $\mathcal{C}$}

\begin{eqnarray}
	\sigma_{x}^{\pm}&\to& (-1)^{x+1}\sigma_{x+1}^{\mp}\\
	S_{x,\hat{i}}^{\pm} &\to& S_{x+1,\hat{i}}^{\mp}
\end{eqnarray}

\emph{The parity transform, $\mathcal{P}$}
\begin{eqnarray}
	\sigma_{x}^{\pm}&\to& \sigma_{-x}^{\pm}\\
	S_{x,\hat{i}}^{\pm} &\to& S_{-(x+1),\hat{i}}^{\mp}
\end{eqnarray}

\begin{figure}[htp]
	\begin{center}
		\includegraphics[width=8.5cm]{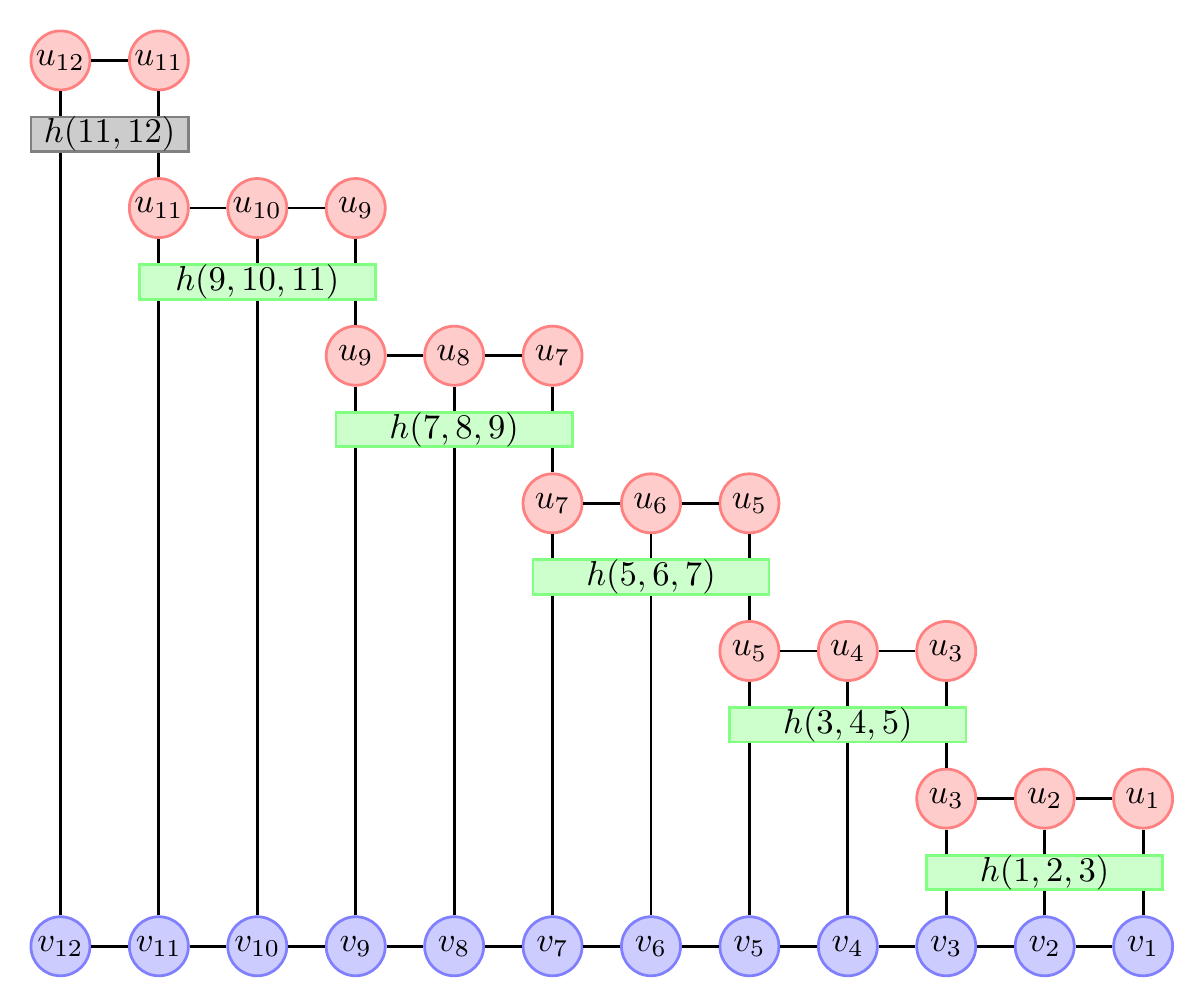}
	\end{center}
	\caption{The scheme for the time-evolution of the wave function. $h(i,j,k)$
	represent the three spin Trotter gate for sites $i,j$ and $k$. The initial
	wave function is represented by blue circles. We act the three spin gates on
	the initial wave function. After the gate application, we decompose the three
	site tensor by singular-value on the two middle links and get the three red
	tensors. At the boundary, we only have two-spin Trotter gate which contains
	only mass term. Then we apply the gates in backward order to finish one
	time-step. We use time step $\delta t=0.02$ and truncation error to be
	$10^{-15}$. The result has no significant change with truncation error $10^{-8}$.}
	\label{fig:sup_0}
\end{figure}

The matrix product state representation is a very efficient way to store many-body quantum
states\cite{Schollwock2011}. 
At equilibrium, the representation has efficient ways to lower the energy of the variational wave
function leads to the success of DMRG which allows us to simulate the quantum states
using a classical computer.
The TEBD algorithm uses the same strategy to store how many-body wave function and evolve
the state in real time\cite{Vidal2003}.
We start with the MPS wave function and evolve the state using Trotter-Suzuki
decomposition of the time-evolution operator as shown in Fig. \ref{fig:sup_0}.

\begin{figure}[htp]
	\begin{center}
		\includegraphics[width=4.5cm]{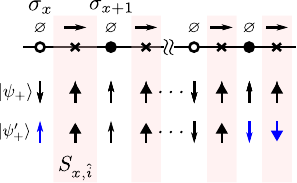}
	\end{center}
	\caption{The wave function with boundary spins modified.}
	\label{fig:sup_1}
\end{figure}

In our simulation, we use open boundary condition. 
In the calculation of Loschmidt echo, instead of using $|\psi_{+}\rangle$ directly, we
use the wave function $|\psi_{+}'\rangle$ where some of the spins at the boundary are
modified as shown in Fig. \ref{fig:sup_1}.
We modify the spin configuration to avoid applying Trotter gate across the system.
Here only the boundary spin configurations are modified, the additional energy cost at
the boundary will be negligible at the thermodynamic limit.

\section{Lanczos-based ED}
\label{app:2d_ED}
Lanczos-based ED is a powerful tool to explore the dynamics of QLM not only because it
exploits the locality of physical Hamiltonian but also
it automatically takes care of the gauge invariance.
The Lanczos-based ED studies the exact dynamics within the Krylov subspace constructed
by successively applying the Hamiltonian to an initial state, $|\phi_i\rangle$.
Because both the Hamiltonian and state are gauge invariant,
the Krylov subspace, $\left\{ |\phi_i\rangle, H|\phi_i\rangle,\cdots,
	H^{M}|\phi_i\rangle \right\}$,
is automatically gauge invariant.
Instead of selecting physical states with a certain winding number and generating
couplings within that sector, the Lanczos-based method automatically selects
relevant states in the sector at runtime.
Taking advantage of the extreme sparse condition, we are able to explore the quench
dynamics through Lanczos-based ED up to a system with $6\times 6$ plaquettes,
equivalent to $72$ spins.
We benchmark our result with $4\times 4$ exact diagonalization where we enumerate all
the gauge invariant states and coupling explicitly. The Lanczos-based ED with $M=20$
reproduces the exact result using full ED in the time window of study. The implementation of Lanczos-based ED can
be found in various articles in the literature~\cite{park1986unitary,sandvik2010}.

\section{long time dynamics at the $m=0$ limit}

\begin{figure}[htp]
	\begin{center}
		\includegraphics[width=8cm]{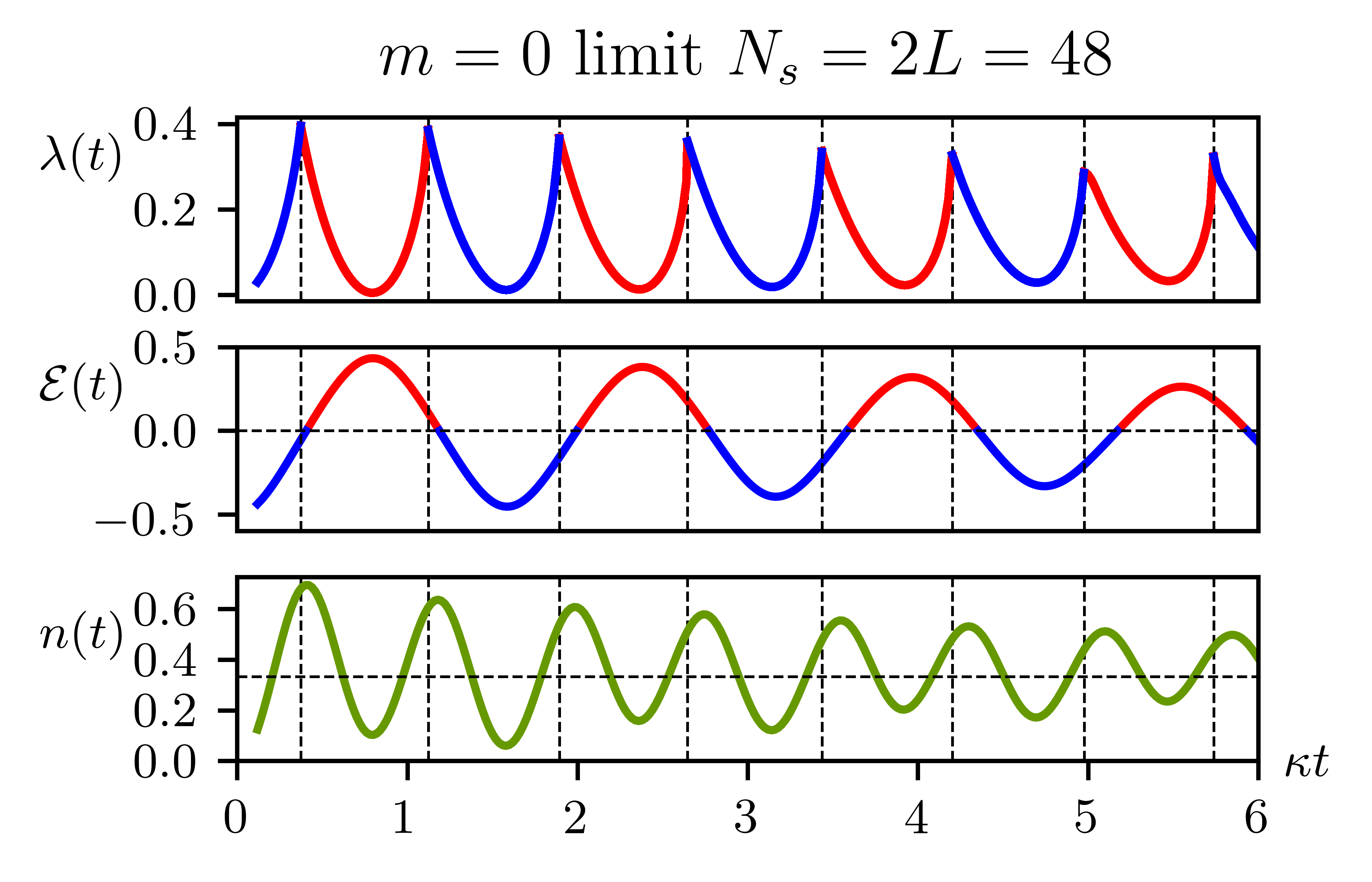}
	\end{center}
	\caption{Long time evolution at the $m=0$ limit.
}
	\label{fig:sup_long_time}
\end{figure}

We explore the time evolution of (1+1)d QLM for system size $L=24$ for longer times
$\kappa t=18$, and at the $m=0$ limit. 
Such choice is equivalent to go to $\kappa_c=0$ limit.
We find a strong boundary effect in the rate function of Loschmidt echo for $\kappa t\gtrapprox 7$. 
We show $\kappa t<6$ in Fig. \ref{fig:sup_long_time}.

We can observe $\mathcal{E}(t)$ oscillating around $\mathcal{E}=0$ as expected since the
system is expected to have an equal probability of $S^z_{x,\hat{i}}=\pm\frac{1}{2}$.
$n(t)$ oscillates near $n=\frac{1}{3}$. 
At long time limit, we expect the system to explore all possibilities. 
A naive estimation for $n(t)$ for long times is $1/2$ since
the probability to be at vacuum and create a particle should be the same.
However, this does not consider Gauss law. The fact that $n(t)$
approaches $1/3$ instead of $1/2$ is because the long time limit can be
considered as the infinite temperature limit where only Gauss law governs the dynamics.
If we consider the constraints are independent for each gauge-particle-gauge unit,
only three configurations can satisfy the Gauss law. That is, $|\downarrow
0 \downarrow \rangle, |\uparrow 0 \uparrow \rangle,
|\downarrow 1 \uparrow \rangle$. Only one of the configuration
has particle excitation which gives the expectation value of $\langle n \rangle=
\frac{1}{3}$. 
The mean value of $\frac{1}{3}$ suggests the role of Gauss law. 
This simple argument, however, assumes every unit is independent, which is not true for the
many-body dynamics.
The precise value for the longtime evolution considering finite size effect requires
more systematic study, which is beyond the scope of this work.

\section{Definition of two component order parameter}
\label{app:2d_OP}
The height variable is an alternative way to denote the flippability of the plaquettes
and is defined on center of the plaquettes.
They take values of $+$ and $-$ and choose the convention that the lower left
plaquette has height $+1$.
The nearby height variable is defined according to the spin configuration on the edge
of the plaquette.
If the spin configuration on the link is pointing toward $+\hat{i},+\hat{j}$
direction, the two plaquettes sharing that edge have opposite heights,
otherwise, the heights stay the same.
The height configurations for the two ground states at $\Lambda=\infty$ are shown
in FIG. \ref{fig:4} (b).
Using the height variable, the two-component order parameter is defined as the height
order in sub-plaquette $A$ and $B$. 
We use the grey background to represent the $A$ sub-plaquette. 
The order parameter is defined as $\hat{M}_{\eta}=N_{P}^{-1}\sum_{i\in\eta}c^{\eta}_i
\hat{h}^{\eta}_i$ where $\hat{h}_{i}^{\eta}$ is the operator that measures the height
variable on plaquette $i$,
$c^{\eta}_i=\langle \phi_{+}|\hat{h}^{\eta}_i|\phi_{+}\rangle$ is the height value
for the plaquette $i$ on sub-plaquette $\eta$ of state $|\phi_{+}\rangle$, and $N_p$
is the total number of plaquettes of the system.
Using this definition the order parameter for $|\phi_{\pm}\rangle$ is
$(M_A,M_B)=\left( \langle \phi_{\pm}|\hat{M}_{A}|\phi_{\pm}\rangle,\langle
\phi_{\pm}|\hat{M}_{B}|\phi_{\pm}\rangle \right)=(0.5,\pm 0.5)$. 

\begin{figure}[htp]
	\begin{center}
		\includegraphics[width=7cm]{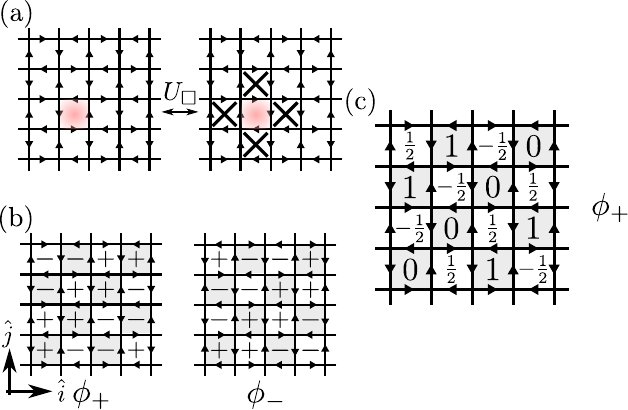}
	\end{center}
	\caption{(a) The tunneling process between two allowed configurations.
	(b) The height representation we use in this work. We use the two ground
	states to represent the phase convention $h(\phi_{+})$ and $h(\phi_{-})$.
	(c) The height representation defined in Ref. \cite{Banerjee2013b}, where we can identify $\left\{ 0,1
	\right\}$ with $\left\{ +,- \right\}$ on sub-plaquette A(grey plaquettes) and
	$\left\{ \frac{1}{2},-\frac{1}{2} \right\}$ with $\left\{ -,+ \right\}$ on
	sub-plaquette B (white plaquettes). As a bookkeeping notation, the definition
	will not change our understanding of the physics.
	}
	\label{fig:sup2}
\end{figure}

\end{document}